%% file: main.tex
\documentclass[10pt,twocolumn,letterpaper]{article}

\usepackage{iccv}              
\usepackage{multirow}
\usepackage[linesnumbered,ruled,vlined]{algorithm2e}
\SetAlgoNoEnd 

\newcommand{\dice}{\texttt{DICE}}


\definecolor{iccvblue}{rgb}{0.21,0.49,0.74}
\usepackage[pagebackref,breaklinks,colorlinks,allcolors=iccvblue]{hyperref}
\usepackage[accsupp]{axessibility}  

\def\paperID{xxx} 
\def\confName{ICCV}
\def\confYear{2025}

\title{DICE: Staleness-Centric Optimizations for Parallel Diffusion MoE Inference}

\author{
Jiajun Luo$^{1*}$, Lizhuo Luo$^{2*}$, Jianru Xu$^{2*}$, Jiajun Song$^1$, Rongwei Lu$^1$, Chen Tang$^3$, Zhi Wang$^{1\dagger}$ \\
$^1$Shenzhen International Graduate School, Tsinghua University \\ \quad $^2$Southern University of Science and Technology \quad $^3$The Chinese University of Hong Kong \\
{\tt\small luo-jj24@mails.tsinghua.edu.cn, wangzhi@sz.tsinghua.edu.cn} 
}

\begin{document}
\twocolumn[{
\renewcommand\twocolumn[1][]{#1}
\maketitle
\begin{center}
    \centering
    \captionsetup{type=figure}
    \vspace{-25pt}
    \includegraphics[width=\textwidth]{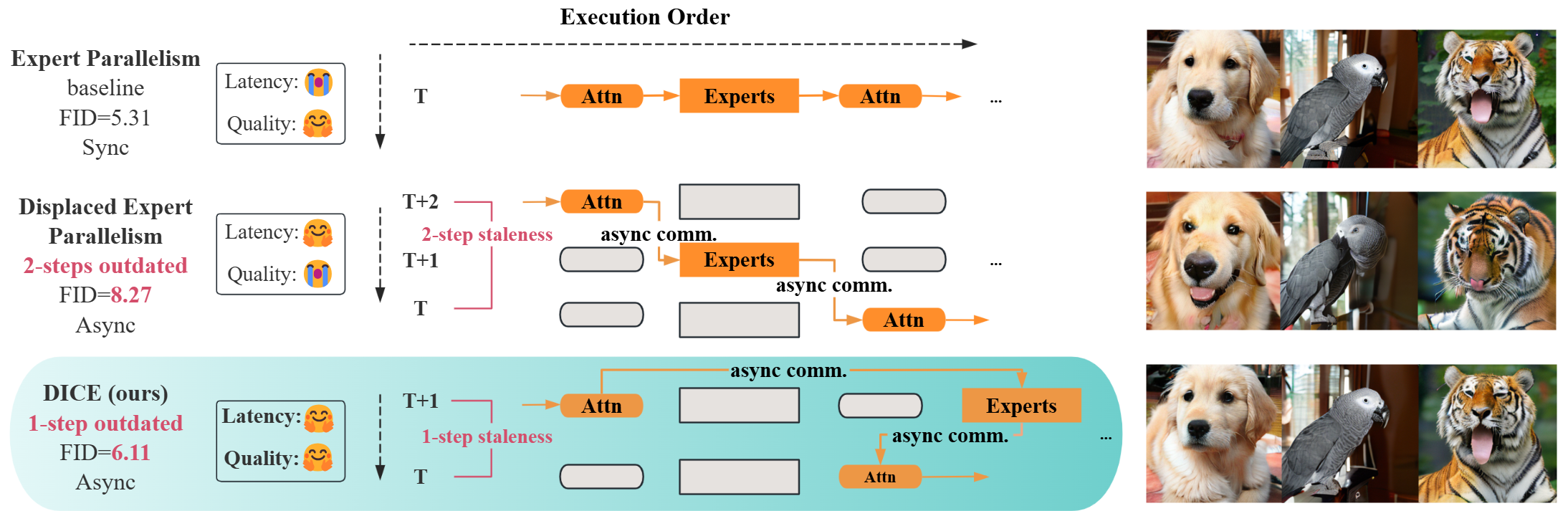}
    \vspace{-22pt}
    \caption{
    We introduce \textbf{\dice{}}, a framework designed to enhance scalability and efficiency in MoE-based diffusion models. Existing asynchronous methods accelerate parallel inference but suffer from outdated activations (staleness). \dice{} tackles this issue with step-, layer-, and token-level optimizations, reducing staleness and achieving up to $1.26\times$ speedup with minimal quality loss. Left: Architectural comparison; Right: Visual quality results.
    }
    \label{fig:teaser}
    \vspace{-5pt}
\end{center}
}]
\maketitle
\def\thefootnote{*}\footnotetext{Equal contribution. Work done when Lizhuo Luo and Jianru Xu were interns at mmlab@SIGS, Tsinghua University.}
\def\thefootnote{{$\dagger$}}\footnotetext{Corresponding author.}
\def\thefootnote{\arabic{footnote}}
\input{sec/abstract}
\input{sec/intro}

\input{sec/related}

\input{sec/motivation}
\input{sec/method}

\input{sec/experiments}
\input{sec/conclusion}

{
    \small
    \bibliographystyle{ieeenat_fullname}
    \bibliography{main}
}

\input{sec/X_suppl}
\end{document}

%% file: sec/abstract.tex
\begin{abstract}

\vspace{-13pt}

Mixture-of-Experts-based (MoE-based) diffusion models demonstrate remarkable scalability in high-fidelity image generation, yet their reliance on expert parallelism introduces critical communication bottlenecks. State-of-the-art methods alleviate such overhead in parallel diffusion inference through computation-communication overlapping, termed displaced parallelism.  However, we identify that these techniques induce severe \textbf{staleness} --- the usage of outdated activations from previous timesteps that significantly degrades quality, especially in expert-parallel scenarios. We tackle this fundamental tension and propose \textbf{\dice{}}, a staleness-centric optimization framework with a three-fold approach: (1) \textit{Interweaved Parallelism} introduces staggered pipelines, effectively halving step-level staleness for free; (2) \textit{Selective Synchronization} operates at layer-level and protects layers vulnerable to staled activations; and (3) \textit{Conditional Communication}, a token-level, training-free method that dynamically adjusts communication frequency based on token importance. Together, these strategies effectively reduce staleness, achieving $1.26\times$ speedup with minimal quality degradation. Empirical results establish \dice{} as an effective and scalable solution. Our code is available at \href{https://github.com/Cobalt-27/DICE}{https://github.com/Cobalt-27/DICE}
\end{abstract}

%% file: sec/intro.tex
\section{Introduction}
\label{sec:intro}

\begin{figure*}[t]
    \centering
    \includegraphics[width=\linewidth]{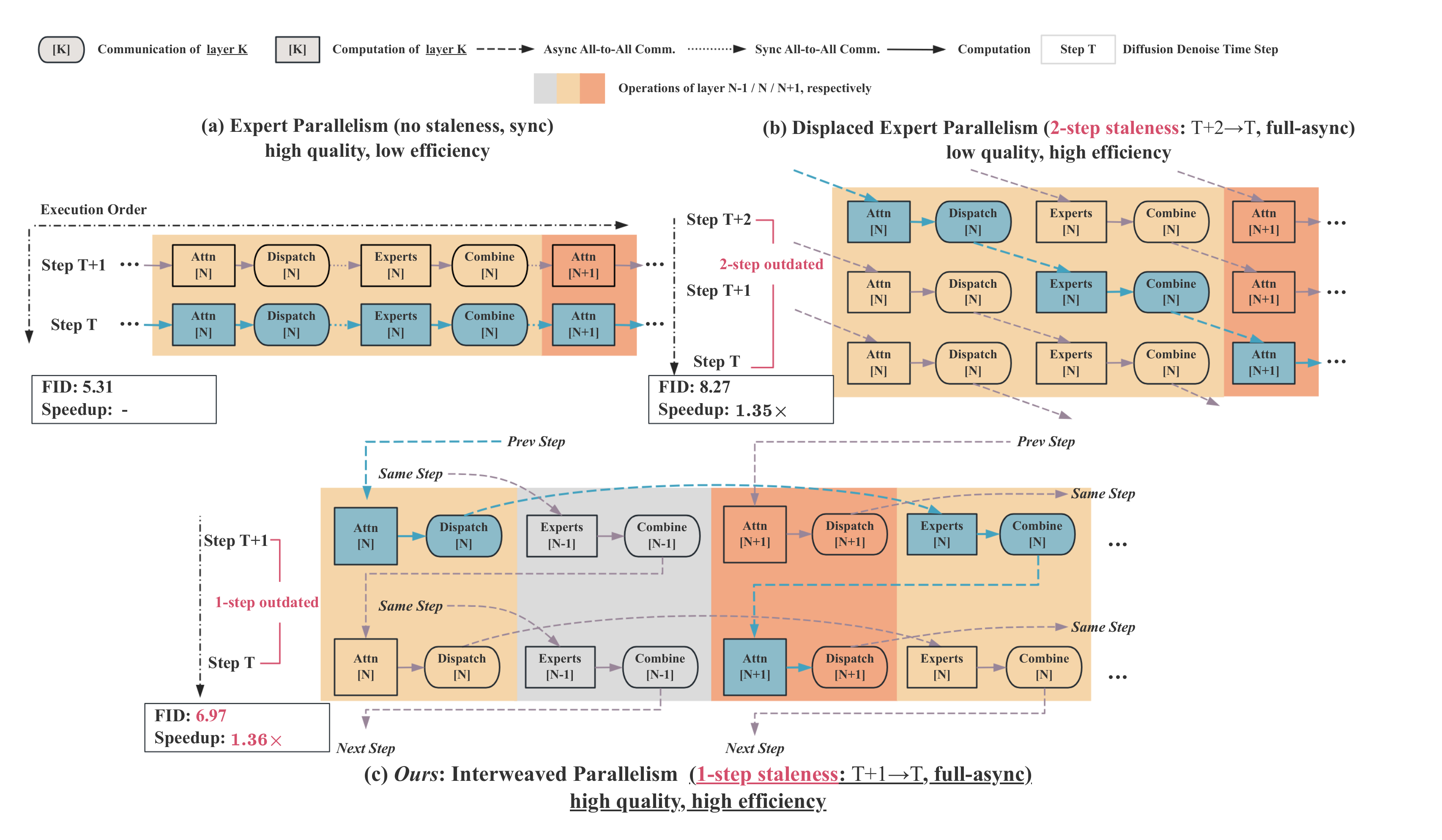}
    \vspace{-20pt}
    \caption{Execution flows of (a) synchronous \textit{expert parallelism} (no staleness); (b) \textit{displaced} variant of \textit{expert parallelism} (two-step staleness, proposed in DistriFusion)~\cite{distri} and (c) our \textit{interweaved parallelism} (one-step staleness). Each color denotes a different layer, showing how interweaving reduces staleness by staggering operations within the same step. The blue path highlights layer~\(N\), which spans two more steps in displaced parallelism but only one in interweaved. FID and speedup refer to interweaved parallelism alone (Section \ref{sec:tradeoff}).}
    \label{fig:para}
    \vspace{-10pt}
\end{figure*}

Diffusion models~\cite{ddpm, dhariwal2021diffusion} have revolutionized the field of generative modeling, enabling the creation of high-fidelity images~\cite{restoration, score} and videos~\cite{video, luo2023videofusion}, rivaling human creativity in quality and detail. Among the architectures, the Diffusion Transformer (DiT)~\cite{dit, transformer, qdit} with Mixture-of-Experts (MoE)~\cite{moe} stands out due to its effectiveness in scalability. MoE enables large diffusion models to scale up with sub-linear cost increases. It divides the model into multiple experts, with only a subset activated for each input. This selective activation reduces computation, achieving high performance at a lower cost. As demonstrated by DiT-MoE~\cite{ditmoe, rap}, MoE-based diffusion models have been scaled to 16 billion parameters with superior quality.

Scaling MoE-based diffusion models necessitates expert parallelism~\cite{gshard, deepspeedmoe}  to handle their extensive memory footprint; however, it also introduces significant communication overhead. By distributing expert weights across multiple GPUs, expert parallelism substantially reduces memory usage but incurs significant communication overhead~\cite{fastermoe, baselayers, interlayer, lina}. As outlined in Figure~\ref{fig:para}(a), two all-to-all communications (dispatch + combine) are required per layer. This bottleneck presents a major challenge for scaling diffusion models.

To address communication challenges in parallel inference, state-of-the-art methods employ \textit{Displaced Parallelism}~\cite{distri}. In standard synchronous execution, communication blocks computation, forcing devices to wait for activations to be transmitted. Displaced parallelism mitigates this bottleneck by leveraging activation similarity across steps: each step sends activations immediately, which are then used in the next step (though outdated). This ``send now, receive next'' strategy prevents blocking while still providing a close approximation of current data due to similarity (also termed redundancy). As visualized in Figure~\ref{fig:para}(b), experts of layer $N$ at step \(T+1\) directly use activations transmitted from step \(T+2\), enabling computation-communication overlap.

However, displaced parallelism introduces the issue of \textbf{staleness}—\textit{the utilization of activations from earlier steps instead of real-time data}—which can significantly degrade model performance, increasing the Fréchet Inception Distance (FID) score from 5.31 to 8.27 (detailed in Section \ref{sec:experiments}). Staleness occurs because asynchronous communication delays the usage of activations until a future step, causing layers to compute based on outdated information. 

We quantify staleness as the \textit{difference in steps between when the input was generated and the step in which its corresponding output is used}. Displaced parallelism exhibits 2-step staleness, as the result used in Step $T$ is driven by activations from Step $T+2$ in the same layer (highlighted in Figure \ref{fig:para}(b)), relying on outdated data. Our method reduces this to 1-step, resulting in improved quality.

We conduct an in-depth analysis of the staleness phenomenon in diffusion parallelism and derive several key insights. At the step level, we find an optimized parallel scheme to halve the staleness compared to displaced parallelism yielding better quality without additional overhead. At the layer level, we show that deeper MoE layers are more sensitive to staleness due to the characteristics of vision tasks and MoE. At the token level, our analysis demonstrates that router-score-driven signals effectively identify critical tokens vulnerable to staleness.

Based on these findings, we propose \textbf{\dice{}} (\textbf{D}iffusion \textbf{I}nference with staleness-\textbf{CE}ntric optimizations), which enhances parallel inference at the step, layer, and token granularities. \dice{} incorporates a three-fold approach to manage staleness effectively: (1) \textit{Interweaved Parallelism} reduces step-level staleness from 2-steps to 1-step compared to displaced expert parallelism, halving the required buffer size, and achieving this without additional overhead. Detailed in Figure \ref{fig:para}(c). (2) \textit{Selective Synchronization} synchronizes only the layers most sensitive to staleness, ensuring that critical information remains up-to-date; and (3) \textit{Conditional Communication}, a fine-grain strategy that adjusts communication frequency based on token importance. Together, these optimizations mitigate staleness issues while enhancing memory usage and inference efficiency, achieving up to a $1.26\times$ speedup with minimal quality impact.

To summarize, our contributions are as follows:

\begin{itemize} 

\item We identify the issue of staleness in MoE-based diffusion model inference, highlighting its impact on performance. 

\item We conduct a comprehensive analysis of staleness, revealing these key insights: optimizing the parallel scheme can reduce staleness for free at the step level; deeper MoE layers are more sensitive to staleness at the layer level, and certain tokens are more adversely affected by staleness at the token level.

\item Based on these insights, we propose \dice{}, which optimizes staleness at the step, layer, and token granularities through interweaved parallelism, selective synchronization, and conditional communication, respectively. \dice{} outperforms state-of-the-art methods, providing a cohesive framework to improve the quality, efficiency, and scalability of parallel inference. Our code is  \href{https://github.com/Cobalt-27/DICE}{publicly available}.

\item We validate our solution across diverse configurations, including different model sizes, hardware and baselines, demonstrating superior efficiency and quality.

\end{itemize}

%% file: sec/related.tex
\section{Preliminaries \& Related Works}
\label{sec:related}

\subsection{MoE-based Diffusion Models}

\noindent Mixture-of-Experts (MoE)~\cite{moe} is an architecture that effectively scales model capacity by partitioning the network into multiple experts, with only a subset activated per input; further detailed in supplement Section~\ref{sec:more-moe}. DiT-MoE \cite{ditmoe} integrates MoE into Diffusion Transformers to effectively scale diffusion models to 16.5 billion parameters. It also employs shared experts to capture common knowledge \cite{deepspeedmoe}. The largest DiT-MoE model features 32 experts and achieves state-of-the-art performance on ImageNet~\cite{imagenet}.

\subsection{Displaced Parallelism}

\noindent Displaced Parallelism, introduced by DistriFusion~\cite{distri}, leverages activation similarity between successive diffusion steps to asynchronously transfer activations, overlapping communication and computation for speedup. Activations computed in the current step are sent asynchronously and used in the next step. This mechanism is expanded upon in supplement Section ~\ref{sec:more-displaced}. However, this approach introduces staleness due to deferred communication, leading to quality degradation, especially in expert parallelism. To address this, we introduce staleness-centric optimizations that enhance efficiency and maintain output quality. PipeFusion~\cite{pipefusion} and AsyncDiff~\cite{asyncdiff} also exploit activation similarity to accelerate diffusion inference. In addition to displaced parallelism, caching is an alternative optimization targeting both U-Net~\cite{deepcache, wimbauer2024cache, habibian2024clockwork, so2023frdiff} and DiT~\cite{fora,learn2cache}.

\subsection{Expert Parallelism}

Expert Parallelism~\cite{gshard} scales MoE models by distributing expert weights across devices while replicating non-expert layers. The input batch is split among devices (similar to data parallelism), and tokens (patches) are routed via all-to-all communication, enabling large-scale deployment. Compared to other parallelism strategies\cite{megatron,lu2023dagc, lu2024tmc, dist_survey}, expert parallelism is inherently more compatible with MoE for its alignment with the architecture.
\begin{figure}[t]
    \centering
    \vspace{-10pt}
    \includegraphics[width=\linewidth]{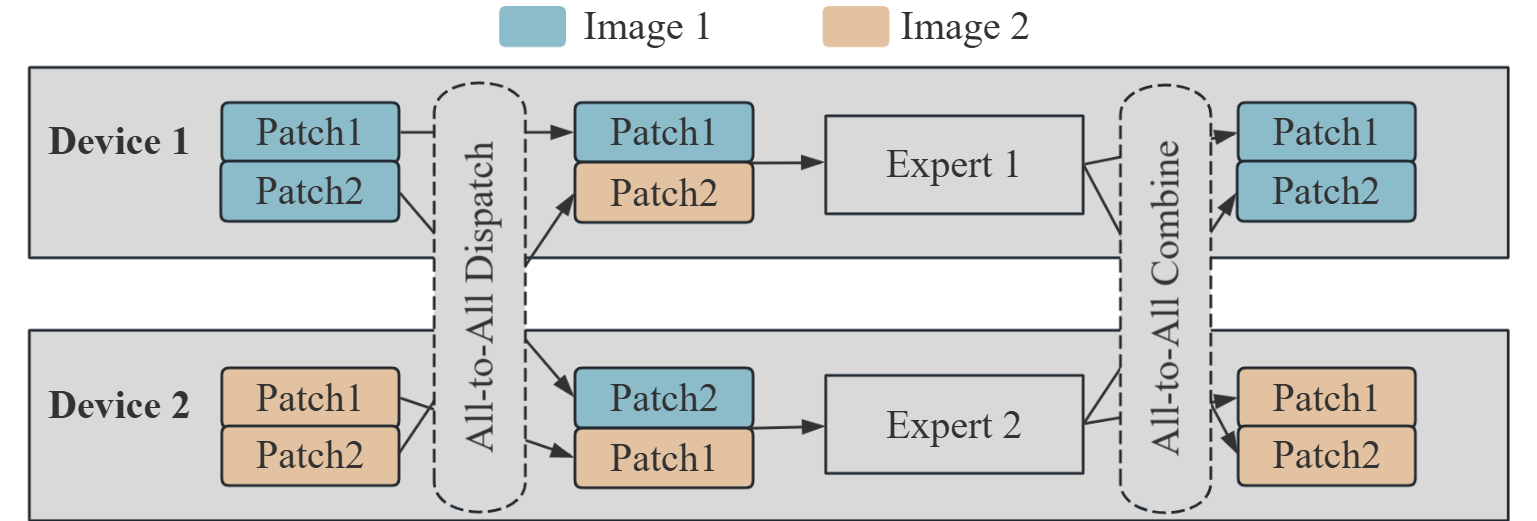}
    \caption{Visualization of expert parallelism. Each device holds a subset of experts (Feed-Forward Networks, FFNs) and processes a portion of the input batch. Different colors denote distinct data samples.}
    \label{fig:ep_intro}
    \vspace{-10pt}
\end{figure}

MoE's strong performance and reliance on expert parallelism have motivated optimizations. FasterMoE~\cite{fastermoe} employs pipelined all-to-all communication and expert shadowing, DeepSpeed-MoE~\cite{deepspeedmoe} leverages efficient communication and model compression; additional work addresses communication via topology-aware token routing~\cite{locmoe, fastermoe}.

However, these methods are not tailored for diffusion models and do not tackle the unique challenges in diffusion processes. Our work fills this gap by introducing staleness-centric optimizations.

%% file: sec/motivation.tex
\section{Motivation}
\label{sec:motivation}

Our motivation is based on three key observations: 
first, expert parallelism suffers from a critical communication bottleneck that necessitates optimization; second, MoE-based diffusion models exhibit high redundancy in their activations and routing, suggesting the feasibility of asynchronous communication; and third, we identify staleness, a direct consequence of such asynchrony, as the core challenge to performance.

\noindent\textbf{Communication bottleneck in expert parallelism.} All-to-all operations in expert parallelism introduce significant communication overhead, presenting a major bottleneck~\cite{fastermoe, baselayers, interlayer, lina}. Our evaluation of DiT-MoE-XL on 8 GPUs shows that all-to-all communication accounts for a substantial portion of the total inference time. Specifically, for batch sizes 4, 8, and 16, the all-to-all communication times were 15.91 seconds (61.7\% of the total time), 28.99 seconds (69.8\%), and 54.94 seconds (73.3\%), respectively, highlighting the necessity to mitigate the communication inefficiency.

\begin{figure}[t]
    \centering
    \includegraphics[width=\linewidth]{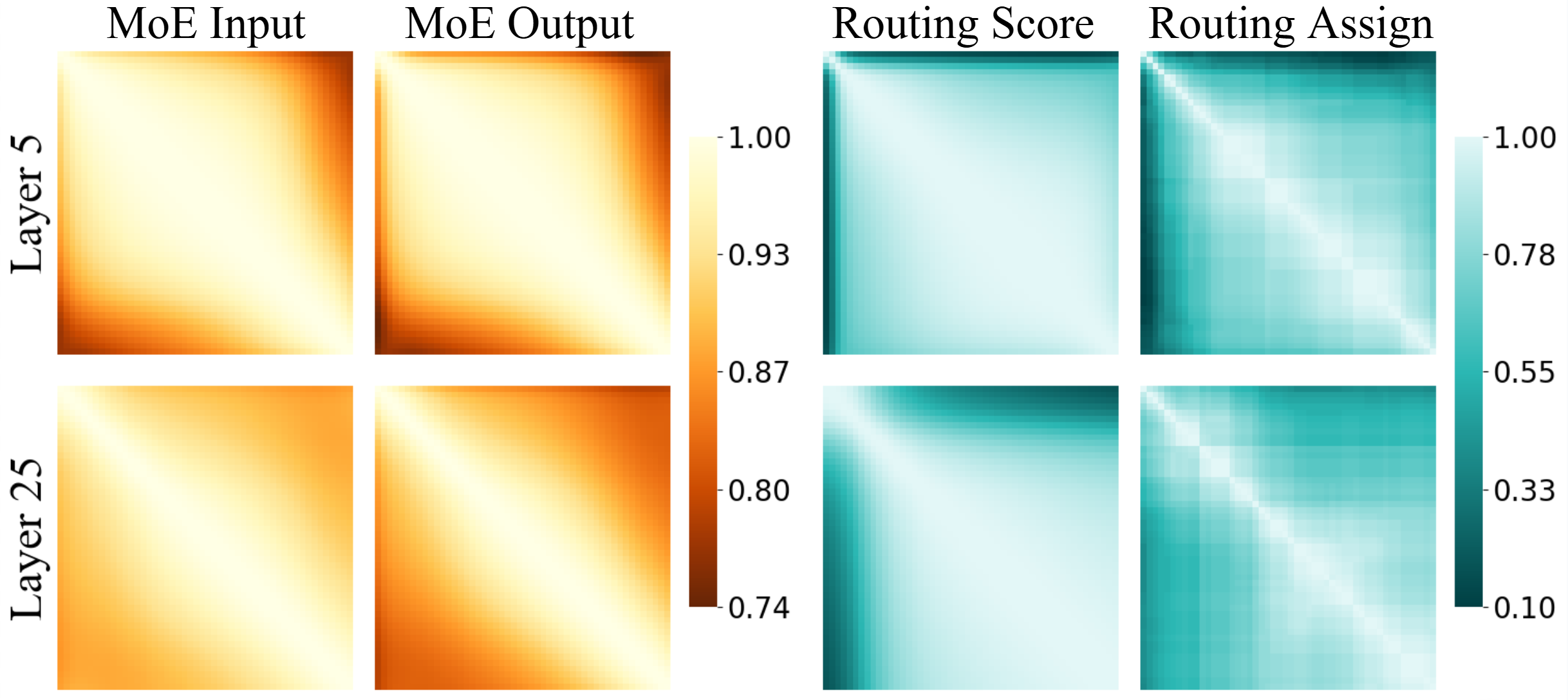}
    \caption{ Step-wise similarity heatmaps (cosine similarity between diffusion steps) in DiT-MoE, with both axes as diffusion steps. Results are shown for Layer 5 and 25, using one-hot routing assignments for similarity computation.} 
    \label{fig:activation_similarity}
\end{figure}

\noindent\textbf{Redundancy in MoE-based Diffusion Models.}
We identify significant computational redundancy in DiT-MoE's expert routing: adjacent diffusion steps exhibit similar token-expert assignments (Figure~\ref{fig:activation_similarity}) and activation values. This strong consistency in both activation ~\cite{deepcache, learn2cache} and routing decisions persists across denoising steps. Such redundancy enables asynchronous communication strategies in expert-parallel systems.

\noindent\textbf{Staleness degrades performance.} We discovered that using outdated (stale) activations due to communication delays in expert parallelism severely degrades model performance. Displaced parallelism induces a two-step staleness in expert parallelism, which causes a drop in image quality: the FID score increases from 5.31 to 8.27 (Section \ref{sec:experiments}). 

%% file: sec/method.tex
\section{Methodology}
\label{sec:method}

\begin{figure*}[t]
    \centering
    \includegraphics[width=\linewidth]{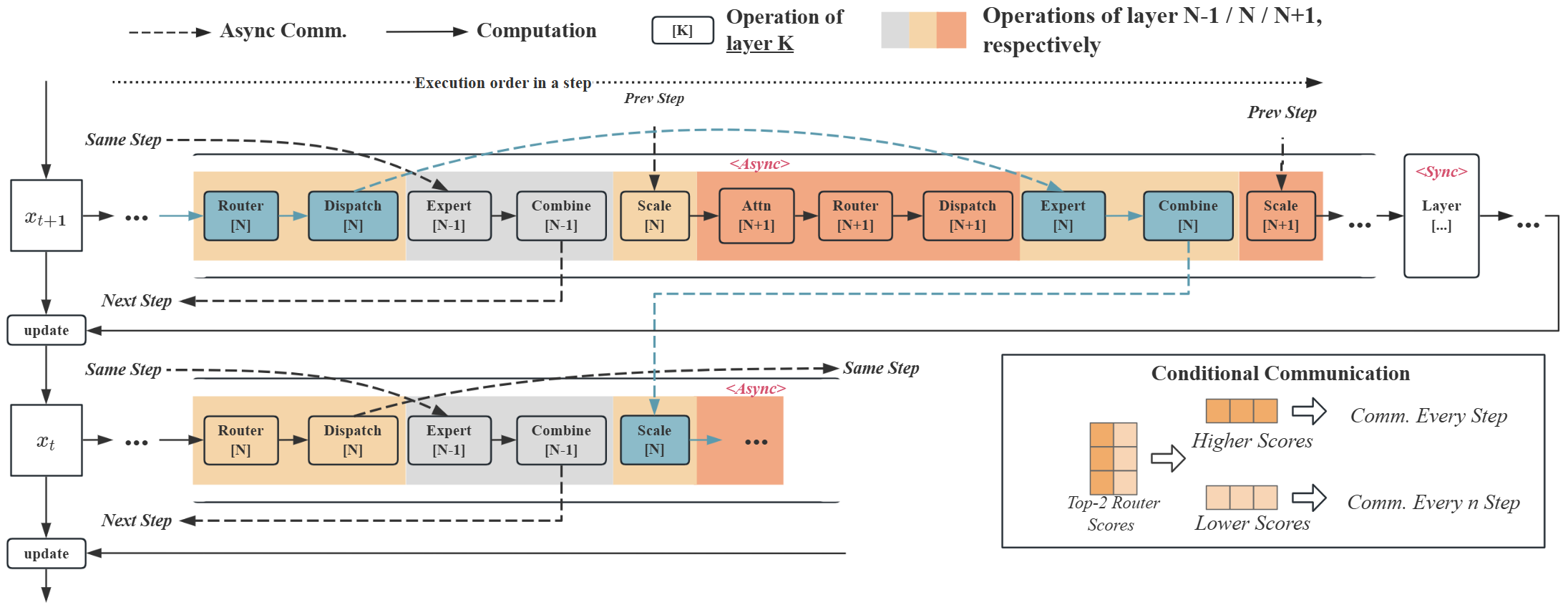}
    \vspace{-10pt}
    \caption{Overview of \dice{}. The blue path highlights layer N’s dataflow. \textit{Interweaved Parallelism} interleaves operations; This execution pattern reducing the steps that cause staleness from two to one, half of displaced expert parallelism \cite{distri}. \textit{Selective Synchronization} targets staleness-vulnerable deeper layers, and \textit{Conditional Communication} prioritizes important tokens based on router scores.}
    \vspace{-5pt}
    \label{fig:arch}
\end{figure*}

Efficient parallel inference for MoE-based diffusion models must address \textbf{staleness}—the use of outdated activations caused by asynchronous communication. While displaced expert parallelism \cite{distri} accelerates inference through overlapping, its asynchronous nature induces severe staleness, compromising output quality.

We propose \textbf{\dice{}}, a staleness-centric solution designed to reduce staleness on step, layer, and token levels. We introduce a three-fold optimization strategy: \textit{Interweaved Parallelism}, \textit{Selective Synchronization}, and \textit{Conditional Communication}. Each component addresses a distinct level of staleness, providing a cohesive framework to boost efficiency and performance.

\subsection{Interweaved Parallelism}
\label{sec:interweaved}

Our first optimization, \textit{Interweaved Parallelism}, redefines the timing of communication and computation to reduce step-level staleness.

In Expert Parallelism, tokens are dispatched via an all-to-all operation to designated experts across devices, processed by local experts, and then returned to their original devices through another all-to-all combine operation. This synchronous process incurs considerable latency, impacting inference speed.

State-of-the-art methods use displaced parallelism~\cite{distri, pipefusion} to overlap communication and computation in distributed diffusion inference. By asynchronously initiating communication in one step and using the results in the subsequent step, it reduces blocking time. 

However, displaced parallelism results in a two-step staleness, causing severe quality degradation: tokens dispatched in one step only reach their designated experts in the next, and the results are then combined two steps later, with FID increased from 5.31 to 8.27.

The proposed \textit{Interweaved Parallelism} halves step-level staleness (2-step $\rightarrow$ 1-step) through improved communication scheduling. As Figure~\ref{fig:para}(b) demonstrates, displaced parallelism introduces 2-step staleness by deferring both all-to-all communication into next step:
$$
\text{Staleness}_{\,\ \text{displaced}} = \underbrace{1}_{\text{dispatch}} + \underbrace{1}_{\text{combine}} = 2\text{-steps}.
$$

Our method achieves 1-step staleness via three key mechanisms: (1) Launching asynchronous dispatch with staggered execution (0-step delay) that completes within the current step, as highlighted in blue in Figure~\ref{fig:arch}, (2) Processing expert outputs while initiating the next combine operation, and (3) Finalizing combined results in the subsequent step (1-step delay). Interweaved parallelism alone improves FID from 8.27 to 6.97. Our solution establishes:
$$
\text{Staleness}_{\,\ \text{interweaved}} = \underbrace{0}_{\text{dispatch}} + \underbrace{1}_{\text{combine}} = 1\text{-step}.
$$

Notably, this scheduling interleaves operations across layers while preserving two critical properties: (i) original dataflow dependencies remain unmodified, and (ii) all computations strictly adhere to their designated layers without cross-layer activation fusion or reuse.

Interweaved parallelism achieves similar overlap to displaced parallelism, as each all-to-all communication still overlaps with one layer’s computation. It also halves the buffer size, storing only the combined results for the next step, unlike displaced parallelism, which buffers both dispatched and combined results.

Interweaved parallelism serves as a free-lunch optimization over displaced parallelism, halving staleness for better image quality, reducing buffer size,  while achieving the same degree of overlap.

\begin{figure}[t]
    \centering
    \vspace{-5pt}
    \includegraphics[width=\linewidth]{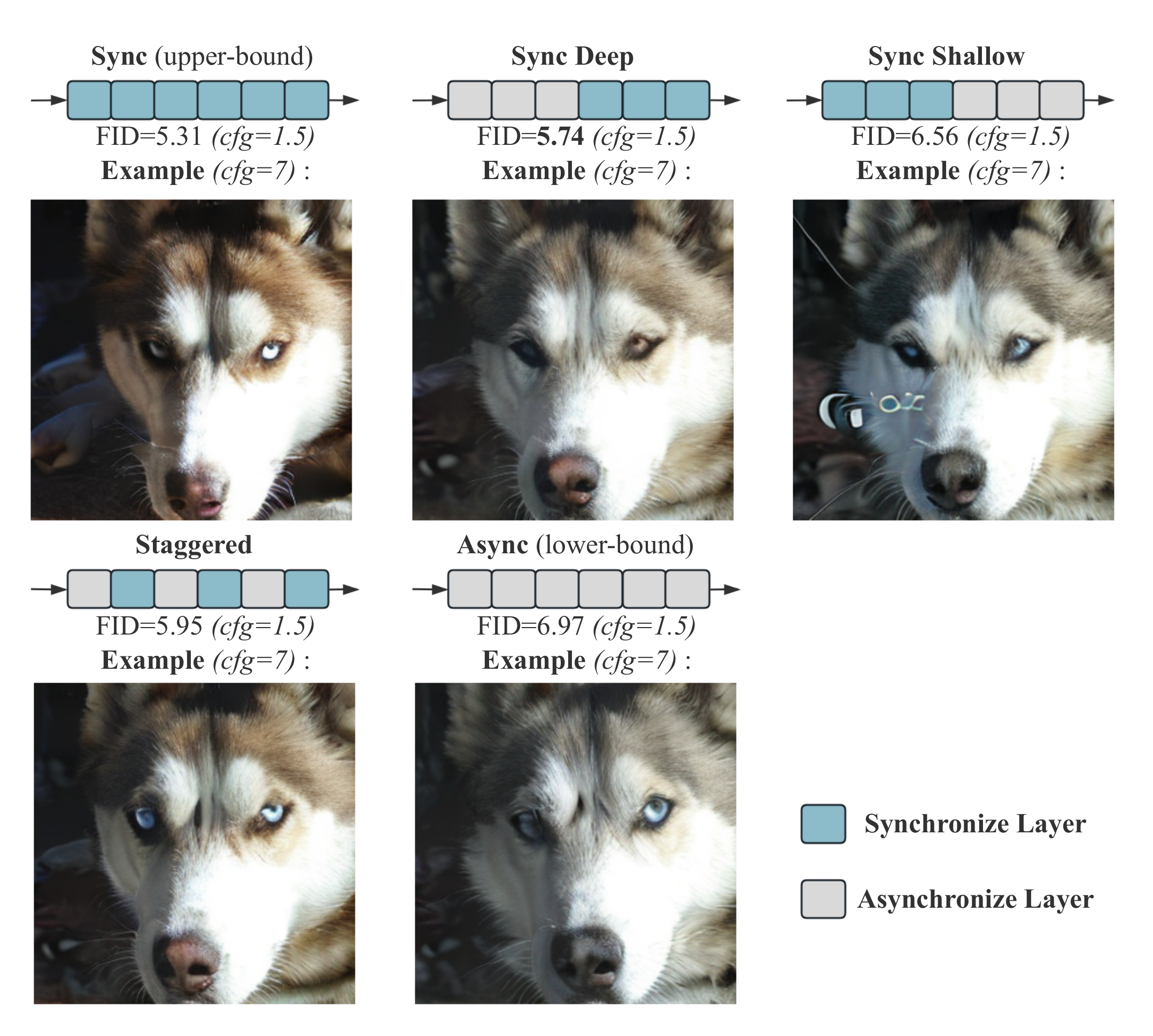}
    \vspace{-25pt}
    \caption{Visual comparison of synchronization strategies in DiT-MoE-XL. Synchronizing only the deep layers provides the most effective optimization. FID shown for \textit{cfg = 1.5}.}
    \vspace{-10pt}
    \label{fig:selective-sync}
\end{figure}

\begin{figure*}[ht]
    \centering
    \includegraphics[width=\linewidth]{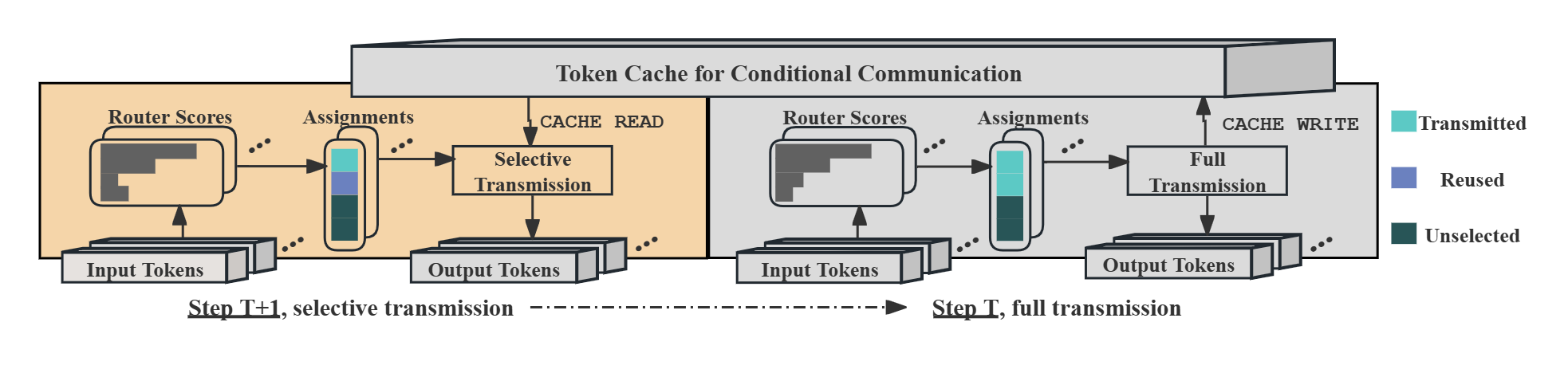}
    \vspace{-25pt}
    \caption{Illustration of \textit{Conditional Communication}. High-scoring tokens are transmitted every step to maintain freshness, while lower-scoring tokens reuse cached value and update every few steps (stride=2 in this example).''Unselected'' refers to scores outside the top-K choices(K is the number of activated experts per token), indicating that the corresponding experts are not activated and no data should be transmitted.}
    \vspace{-5 pt}
    \label{fig:cond-comm}
\end{figure*}

\subsection{Selectively Synchronize Vulnerable Layers} 
\label{sec:selective}

At the layer level, selectively synchronizing vulnerable layers can significantly improve image quality. Our analysis reveals that the impact of staleness exhibits layer-wise asymmetry in MoE-based diffusion models: shallow experts extracting low-level features are inherently robust to asynchronous communication, while experts in deeper layers handling high-level semantics prove fundamentally vulnerable to activation staleness. This architectural property aligns with DeepSpeed-MoE's observation in language models~\cite{deepspeedmoe} where deeper layers benefit more from MoE.

Guided by this understanding, we propose \textit{Selective Synchronization}, which synchronizes only the more vulnerable deeper layers while allowing shallow layers to continue asynchronously. This targeted approach ensures that deeper layers benefit from up-to-date information without the need for full synchronization. Experimental validation in Figure~\ref{fig:selective-sync} confirms the layer-wise vulnerability pattern. Ablation studies (Section \ref{sec:ablation}) demonstrate how our partial synchronization achieves optimal quality-efficiency tradeoffs.

\subsection{Freshness-Latency Trade-Off via Conditional Communication} 
\label{sec:cond}
Balancing freshness and communication efficiency requires token-level prioritization (Figure~\ref{fig:arch}). We achieve this by exploiting inherent properties of MoE routing in diffusion models: (1) High-score tokens dominate output through router score-weighted summation. Thus, more vulnerable to staleness perturbation (Equation~\ref{eq:grad}); (2) Their expert assignments exhibit stronger semantic alignment, making stale activations more detrimental. Our solution-\textit{Conditional Communication}, dynamically adjusts communication frequency through score-driven signals.

To quantify staleness impact, consider a token $i$ assigned to expert $e$ with expert activations $\mathbf{h}_i^e$ and router scores $s_i^e$, the output $\mathbf{y}_i = \sum_{e=1}^E s_i^e\mathbf{h}_i^e$ with L2-norm $\|\mathbf{y}_i\|$. The gradient:
\begin{equation}
\label{eq:grad}
\frac{\partial \|\mathbf{y}_i\|}{\partial \mathbf{h}_i^e} = \frac{s_i^e \cdot \mathbf{y}_i}{\|\mathbf{y}_i\|},
\end{equation}
implies staleness-induced activation perturbations ($\Delta\mathbf{h}_i^e$) propagate to outputs proportionally to $s_i^e$.

Our training-free solution maintains freshness for critical tokens while tolerating staleness in less impactful ones, as shown in Figure~\ref{fig:cond-comm}; elaborated in supplement Section~\ref{sec:more-impl}. It achieves substantial communication reduction with minimal quality trade-off, as validated in Section~\ref{sec:ablation}.

%% file: sec/experiments.tex
\section{Experiments}
\label{sec:experiments}

\begin{figure*}[h]
    \centering
    \vspace{15pt}
    \includegraphics[width=\textwidth]{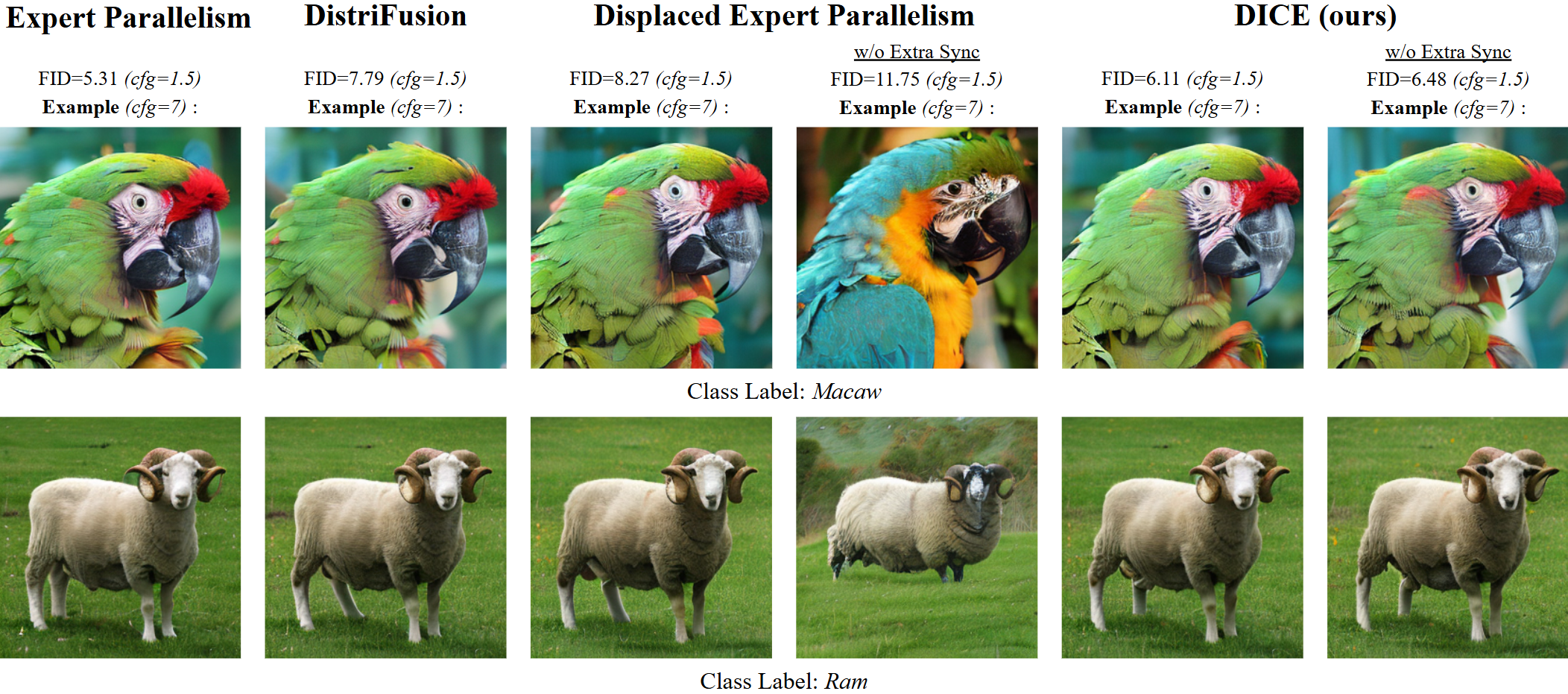}
    \vspace{-20pt}
    \caption{Qualitative results. \dice{} can reduce latency, increase throughput, save memory, and ensure image quality.} 
    \label{fig:img_comp}
\end{figure*}

\subsection{Setups}
\noindent\textbf{Hardware.} Our experiments were conducted on 8 \texttt{NVIDIA GeForce RTX 4090 (24GB)} GPUs connected via \texttt{PCIe}, powered by \texttt{128 vCPU Intel(R) Xeon(R) Gold 6430} CPU.


\noindent\textbf{Models and Datasets.} We evaluate our approach on two configurations of the DiT-MoE model \cite{ditmoe-hf}: DiT-MoE-XL, which is configured with 8 experts across 28 layers, and DiT-MoE-G, with 16 experts across 40 layers, both containing 2 extra shared experts. DiT-MoE is trained on ImageNet, which we use as the benchmark for evaluation.

\noindent\textbf{Baselines.} To evaluate the effectiveness of our approach, we compare it against three baselines: \textit{DistriFusion}~\cite{distri} (displaced sequence parallelism), a state-of-the-art method for distributed diffusion models employing patch parallelism; \textit{Expert Parallelism}~\cite{gshard}, the standard approach for parallelizing MoE models; and \textit{displaced expert parallelism}, the displaced variant~\cite{distri} of expert parallelism, which overlaps communication and computation.

Our main results use a 50-step configuration, which is hardcoded in the DiT-MoE codebase, for fair comparisons. We also provide additional results for 10-and 20-step settings, showing that \dice{} is even more effective under fewer steps.

\noindent\textbf{Metrics.} We assess the performance of our method using several well-established metrics, including Fréchet Inception Distance (FID)~\cite{fid}, Sliced Fréchet Inception Distance (sFID)~\cite{sfid}, Inception Score (IS)~\cite{is}, as well as Precision and Recall~\cite{prec-recall}. 

\noindent\textbf{Implementation Details.} Our implementation is based on \texttt{PyTorch 2.0.0+cu118}~\cite{torch}. We utilize Rectified Flow~\cite{rf,cfg} for the DiT-MoE-XL and G models, as they are specifically trained to support this approach. The core codebase builds upon the original DiT-MoE implementation~\cite{ditmoe-git}. For expert parallelism, we referenced FastMoE~\cite{fastmoe}. Sequence Parallelism~\cite{seqpara} and  DistriFusion~\cite{distri} are adapted to DiT-MoE. We provide more details in supplement Section~\ref{sec:more-impl}.

\begin{table}[t]
    \centering
    \renewcommand{\arraystretch}{1.2}
    \setlength{\tabcolsep}{5pt}
    \resizebox{\columnwidth}{!}{
        \begin{tabular}{lccccc}
            \toprule
            \multicolumn{6}{l}{\textbf{Class-Conditional ImageNet 256$\times$256}} \\
            \midrule
            Method & FID $\downarrow$ & sFID $\downarrow$ & IS $\uparrow$ & Precision $\uparrow$ & Recall $\uparrow$ \\
            \midrule
            Expert Parallelism & 5.31 & 10.10 & 235.89 & 0.75 & 0.60 \\
            \midrule
            DistriFusion & 7.79 & 12.13 & 206.24 & 0.72 & \textbf{0.59} \\
            Displaced Expert Parallelism & 8.27 & 11.58 & 204.07 & 0.71 & \textbf{0.59} \\
            Interweaved Parallelism & 6.97  & 11.01  & 216.62  & 0.72 & \textbf{0.59}\\
            \dice{} &\textbf{6.11} & \textbf{10.93}  & \textbf{225.65} & \textbf{0.73} & \textbf{0.59}\\
            \bottomrule
        \end{tabular}
    }
    \caption{Quantitative evaluation. We employed Rectified Flow with 50 steps to generate 50K samples, evaluating on the ImageNet 256×256 dataset.}
    \vspace{-10pt}
    \label{tab:quality_results}
\end{table}

\begin{figure*}[h]
    \centering
    \vspace{5pt}
    \includegraphics[width=\textwidth]{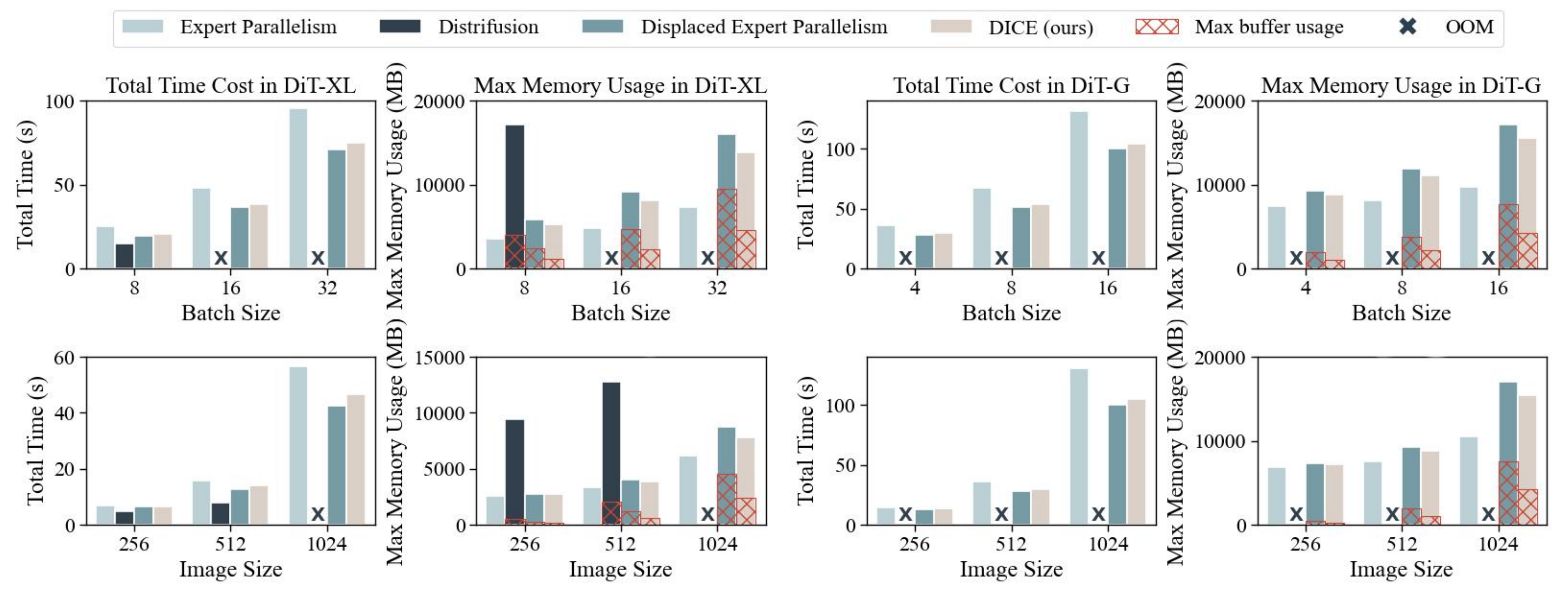}
    \vspace{-20pt}
    \caption{\textbf{Batch size scaling} (at $256\times256$ resolution) and \textbf{Image size scaling} (with batch size 1 per device) performance of DiT-MoE on 8 NVIDIA RTX 4090 GPUs, demonstrating \dice{}'s superior efficiency in speed and memory usage.}
    \label{fig:All}
    \vspace{0 pt}
\end{figure*}

\subsection{Main Results}

\noindent\textbf{Quality results.} In Figure \ref{fig:img_comp}, we present qualitative results and quantitative evaluations in Table \ref{tab:quality_results}. Images generated by \dice{} closely match those from expert parallelism, maintaining high visual fidelity.

\dice{} significantly outperforms DistriFusion and displaced expert parallelism in all quality metrics, notably demonstrating a significantly lower FID of \textbf{6.11} compared to 7.79 (DistriFusion) and 8.27 (displaced expert para.). In fact, interweaved parallelism alone, due to its effective mitigation of staleness, already achieves superior image quality compared to other methods, while \dice{} further increases quality through well-optimized trade-off. 

\noindent\textbf{Speedups and memory analysis.} \dice{} effectively achieves a 1.2$\times$ acceleration over expert parallelism while maintaining the quality. As demonstrated in Figure \ref{fig:All}, \dice{}  consistently exhibits significant speedups over expert parallelism across various batch sizes and image resolutions on both DiT-MoE-XL and G, reaching a maximum of \textbf{1.26$\times$} at batch size 32. 

\dice{} consumes relatively less memory compared to other acceleration methods. Our method requires only half the buffer size of displaced parallelism. However, the mismatch between peak memory and buffer sizes makes the memory optimization less visible in the figure. Compared to DistriFusion, \dice{} has better memory efficiency, as DistriFusion runs out of memory on the XL model at batch size 16 and higher. Additionally, the large parameter size of the DiT-MoE-G model (around 33GB) prevents DistriFusion from running in our setup, highlighting the necessity of expert parallelism in reducing memory footprint.

\begin{table}[t]
    \centering
    \renewcommand{\arraystretch}{1.2}
    \setlength{\tabcolsep}{5pt}
    \resizebox{\columnwidth}{!}{
        \begin{tabular}{lccccc}
            \toprule
            \multicolumn{6}{l}{\textbf{Class-Conditional ImageNet 256$\times$256}} \\
            \midrule
            Method & FID $\downarrow$ & sFID $\downarrow$ & IS $\uparrow$ & Precision $\uparrow$ & Speedup $\uparrow$ \\
            \midrule
            Expert Parallelism & 10.24 & 11.79 & 197.16 & 0.70 & - \\
            \midrule
            DistriFusion & 27.36 & 22.26 & 113.64 & 0.58 & OOM\\
            Displaced Expert Parallelism & 27.61 & 15.42 & 111.05 & 0.55 &\textbf{1.28$\times$} \\
            \dice{} & \textbf{15.13} & \textbf{14.52}  & \textbf{166.32} & \textbf{0.66} & 1.20$\times$ \\
            \bottomrule
        \end{tabular}
    }
    \caption{Results on 10 Steps. All asynchronous methods apply 2 synchronized steps post cold start.}
    \vspace{0pt}
    \label{tab:10step_results}
\end{table}

\begin{table}[t]
    \centering
    \renewcommand{\arraystretch}{1.2}
    \setlength{\tabcolsep}{5pt}
    \resizebox{\columnwidth}{!}{
        \begin{tabular}{lccccc}
            \toprule
            \multicolumn{6}{l}{\textbf{Class-Conditional ImageNet 256$\times$256}} \\
            \midrule
            Method & FID $\downarrow$ & sFID $\downarrow$ & IS $\uparrow$ & Precision $\uparrow$ & Speedup $\uparrow$ \\
            \midrule
            Expert Parallelism & 6.41 & 10.66 & 226.01 & 0.74 & - \\
            \midrule
            DistriFusion & 13.72 & 16.00 & 164.98 & 0.67 & OOM\\
            Displaced Expert Parallelism & 15.27 & 13.44 & 156.82 & 0.64 & \textbf{1.33$\times$}\\
            \dice{} & \textbf{8.60} & \textbf{12.3}  & \textbf{205.05} & \textbf{0.71} & 1.24$\times$ \\
            \bottomrule
        \end{tabular}
    }
    \caption{Results on 20 Steps. We employed Rectified Flow with 20 steps with 4 synchronized steps.}
    \vspace{0 pt}
    \label{tab:20step_results}
\end{table}

\subsection{Experiments on Fewer Steps}
To ensure broader applicability, we performed additional comparative experiments using \textbf{10} and \textbf{20} sampling steps.

\noindent\textbf{Quality results.} Table \ref{tab:10step_results} and \ref{tab:20step_results} show that even in fewer steps, \dice{} continues to exhibit better performance. In terms of quality, \dice{} achieves FID scores of 15.13 and 8.60, which are lower than those of DistriFusion and the displaced expert parallelism. These results further demonstrate that our method achieves acceleration while incurring minimal quality loss. 

\noindent\textbf{Speedups and memory analysis.} \dice{} continues to gain favorable acceleration at reduced steps. As evidenced in Tables \ref{tab:10step_results} and \ref{tab:20step_results}, it maintains a speedup exceeding $1.2\times$. In comparison to DistriFusion, \dice{} consistently exhibits a notable advantage in memory efficiency, whereas DistriFusion encounters out-of-memory (OOM) in these setups.

\begin{table}[t]
    \centering
    \renewcommand{\arraystretch}{1.2}
    \setlength{\tabcolsep}{4pt}
    \resizebox{\columnwidth}{!}{
        \begin{tabular}{cccccc}
            \toprule
            \multicolumn{6}{l}{\textbf{Class-Conditional ImageNet 256$\times$256}} \\
            \midrule
             Interweaved & Selective Sync & Conditional Comm & FID $\downarrow$ & sFID $\downarrow$ & IS $\uparrow$ \\
            \midrule
            \checkmark & \texttimes & \texttimes & 6.97 & 11.01 & 216.62  \\
            \midrule
            \checkmark & \textbf{Deep} & \texttimes & \textbf{5.74} & 10.53 & \textbf{230.23}  \\
            \checkmark & Shallow & \texttimes & 6.55 & 10.63 & 221.61  \\
            \checkmark & Staggered & \texttimes & 5.95 & \textbf{10.39} & 227.78 \\
            \midrule
            \checkmark & \texttimes & \textbf{Low Score} & \textbf{7.24} & \textbf{11.26} & \textbf{214.10}  \\
            \checkmark & \texttimes & High Score & 7.51 & 11.51 & 211.40 \\
            \checkmark & \texttimes & Random & 7.37 & 11.38 & 212.84 \\
            \bottomrule
        \end{tabular}
    }
    \caption{Ablation quantitative evaluation. We applied various strategies to \dice{} to generate 50K samples, evaluated on the ImageNet 256×256 dataset. The rows sequentially compare different strategies for \textbf{selective synchronization} and \textbf{conditional communication}.}
    \vspace{-7pt}
    \label{tab:ablation_quality_results}
\end{table}

\subsection{Ablation Study}
\label{sec:ablation}

\noindent\textbf{Selective Synchronization.}  
Our analysis indicated that deeper layers are more sensitive to staleness. Ablation results (Table \ref{tab:ablation_quality_results}) confirm that synchronizing only the deeper half (\textit{Deep}) yields the best image quality, effectively trading a small latency increase for substantial gains.

\noindent\textbf{Conditional Communication.} 
Our approach leverages router scores as a reliable signal for token importance, updating high-score tokens every step while reducing the communication frequency for lower-score tokens. Ablation results (Table~\ref{tab:ablation_quality_results}) show that deprioritizing lower-score tokens leads to superior image quality, validating the effectiveness and efficiency of our trade-off between quality and reduced communication.

\subsection{Latency-Quality Trade-Off}
\label{sec:tradeoff}

\begin{figure}[t]
    \centering
    \vspace{-10pt}
    \includegraphics[width=\linewidth]{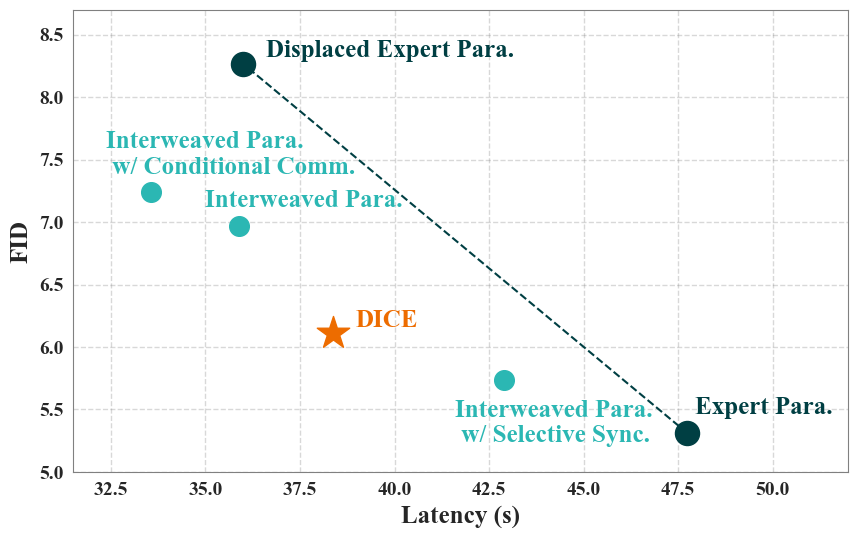}
    \vspace{-20pt}
    \caption{Latency-quality trade-off of proposed optimizations. Interweaved parallelism reduces latency without sacrificing quality, while additional techniques enhance efficiency. Baselines are shown in dark colors, with latency measured at batch size 16. DistriFusion is out-of-memory thus not plotted.}
    \vspace{-10pt}
    \label{fig:trade_off}
\end{figure}

\dice{} achieves an efficient trade-off between quality and latency, as presented in Figure \ref{fig:trade_off}. Interweaved parallelism offers a free-lunch optimization, improving quality with no latency added, while additional techniques further improve efficiency.

%% file: sec/conclusion.tex
\section{Conclusion}
\label{sec:conclusion}
We propose \dice{}, a staleness-centric optimization framework that accelerates MoE-based diffusion model inference through three core innovations: \textit{Interweaved Parallelism} for reducing step-level staleness, \textit{Selective Synchronization} for selectively synchronizing staleness-sensitive layers, and \textit{Conditional Communication}, a token-aware strategy that dynamically adjusts communication frequency. Together, these techniques reduce memory usage, enhance inference efficiency, and achieve up to $1.26\times$ speedup over expert parallelism with minimal quality loss. We anticipate that \dice{} will pave the way for scalability and efficiency improvements in diffusion model serving.

\section*{Acknowledgment}

This work is supported in part by the National Key Research and Development Project of China (Grant No.~2023YFF0905502), National Natural Science Foundation of China (Grant No.~92467204 and 62472249), and Shenzhen Science and Technology Program (Grant No.~JCYJ20220818101014030 and KJZD20240903102300001).

%% file: sec/X_suppl.tex
\clearpage
\setcounter{page}{1}
\maketitlesupplementary

\section{More Preliminaries}
\subsection{Mixture-of-Experts.} 
\label{sec:more-moe}
\begin{figure}[h]
    \centering
    \includegraphics[width=\linewidth]{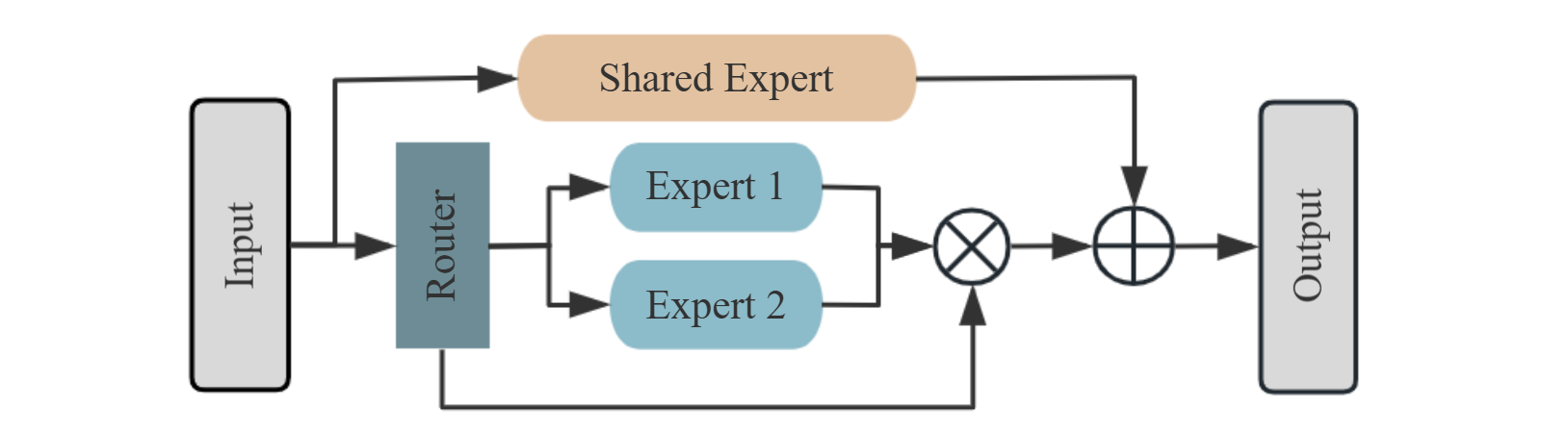}
    \caption{Illustration of an MoE model\cite{ditmoe} with one shared experts and two non-shared experts.}
    \label{fig:moe}
\end{figure}

Figure~\ref{fig:moe} shows the architecture of MoE model. The router directs each token to a subset of experts based on router scores, while a shared expert captures common representations. Outputs from the activated experts are then combined to form the final output. This design allows sublinear scaling by selectively activating only a few experts per input.

\subsection{Displaced Parallelism.} 
\label{sec:more-displaced}

\begin{figure}[h]
    \centering
    \includegraphics[width=\linewidth]{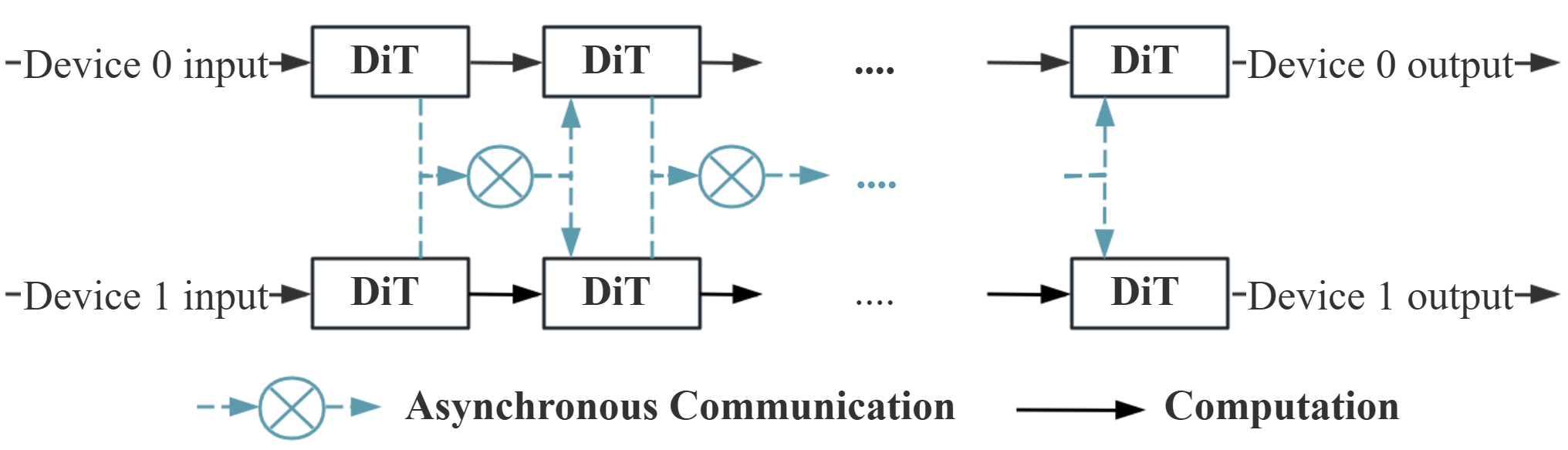}
    \caption{Illustration of displaced parallelism in DiT across multiple devices, with dashed arrows representing asynchronous communication steps that defer data exchange until the next computation stage.}
    \label{fig:df}
\end{figure}

Displaced parallelism~\cite{distri} overlaps communication and computation by asynchronously transmitting activations computed in the current step for use in the next step (Algorithm~\ref{alg:disp_para}).  this prevents blocking (Algorithm~\ref{alg:expert_para}). As shown in Figure~\ref{fig:df}, each device sends activations without waiting, continuing computation with slightly outdated activations from the previous step. This approach prevents communication-induced blocking but introduces staleness, as computations rely on slightly outdated activations rather than fresh data.

\begin{algorithm}[t]
\caption{Expert Parallelism}\label{alg:expert_para}
\For{each layer}{
  \tcp{Perform attention}
  $x \gets \text{attn}(x)$ \\
  $[s,\, idx] \gets \text{router}(x)$ \tcp{Compute router scores and assign tokens to experts}
  $x_d \gets \text{all2all\_dispatch}(x, idx)$ \tcp{Synchronous dispatch (blocking)}
  $x_e \gets \text{expert}(x_d)$ \tcp{Process tokens with local experts}
  $x_c \gets \text{all2all\_combine}(x_e)$ \tcp{Synchronous combine (blocking)}
  $x \gets \text{scale}(x_c, s)$ \tcp{Scale outputs using router scores}
}
\end{algorithm}

\begin{algorithm}[t]
\caption{Displaced Expert Parallelism}\label{alg:disp_para}
\For{each layer}{
  $x \gets \text{attn}(x)$ \\
  $[s,\, idx] \gets \text{router}(x)$ \\
  $h_d \gets \text{async\_dispatch}(x, idx)$ \tcp{Launch non-blocking dispatch (returns handle $h_d$)}
  $x_d \gets \text{wait}(h_d^{prev})$ \tcp{Wait for previous step's dispatch result}
  $x_e \gets \text{expert}(x_d)$ \tcp{Process outdated tokens with local experts}
  $h_c \gets \text{async\_combine}(x_e)$ \tcp{Launch non-blocking combine (returns handle $h_c$)}
  $x_c \gets \text{wait}(h_c^{prev})$ \tcp{Wait for previous step's combine result} 
  $x \gets \text{scale}(x_c, s)$
}
\end{algorithm}

\begin{figure*}[ht]
    \centering
    \includegraphics[width=\linewidth]{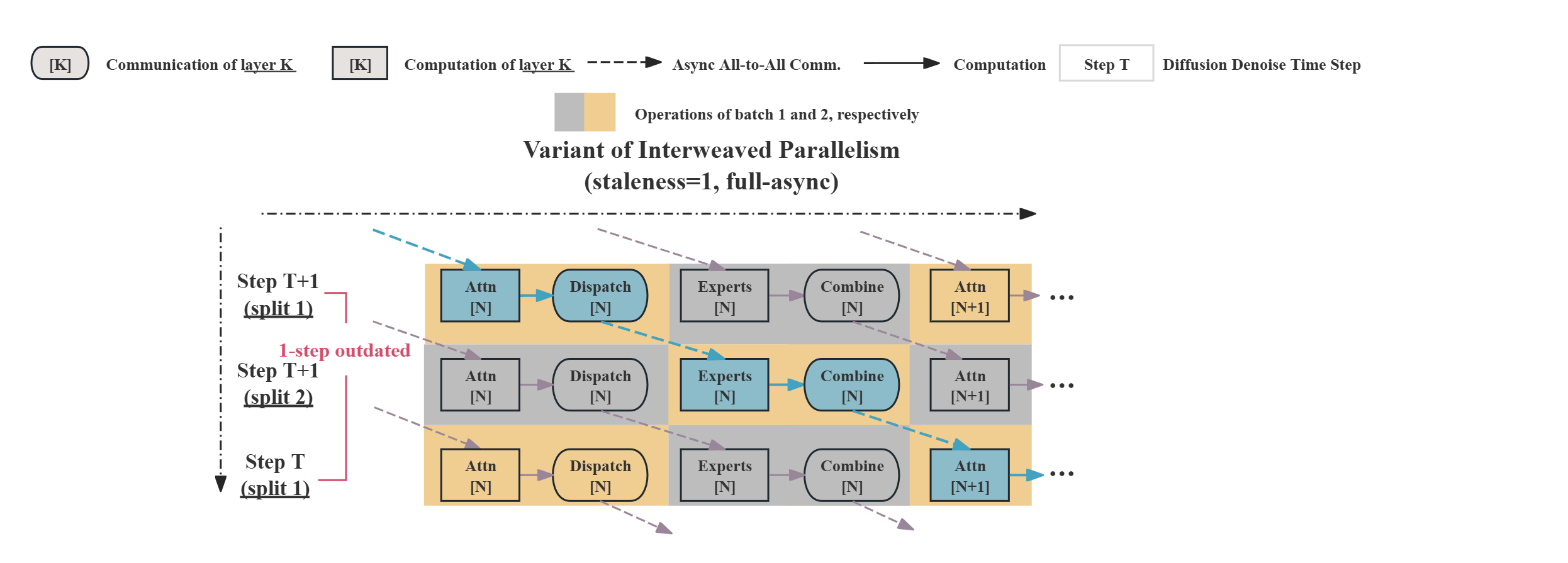}
    \vspace{-25pt}
    \caption{Illustration to the \textit{staggered batch solution}, where local batches are split and interleaved to reduce staleness. However, this approach requires additional buffers, doubles memory usage, and reduces GPU utilization due to smaller effective batch sizes.}
    \vspace{0pt}
    \label{fig:split-batch}
\end{figure*}

\section{Alternative Staggered Batch Solution}
\label{sec:staggered_batch}

Many works split batches to better overlap communication and computation~\cite{fastermoe, lina}. We initially explored a \textbf{staggered batch solution}, in which each device divides its local batch into multiple sub-batches and processes them in a staggered manner. As shown in Figure~\ref{fig:split-batch}, this approach can reduce staleness by handling multiple sub-batches.

However, we ultimately do not adopt this solution for three main reasons:
\begin{itemize}
    \item \textbf{Reduced GPU Utilization.} Splitting the batch lowers the effective batch size per device, potentially underutilizing GPU resources and degrading throughput.
    \item \textbf{Increased Memory Requirement.} Unlike interweaved parallelism—which requires a persistent buffer only for combine and a temporary one for dispatch—the staggered batch solution must maintain persistent buffers for both dispatch and combine, effectively doubling memory usage and increasing the risk of out-of-memory.
    \item \textbf{Batch Size Limitations.} This solution requires a local batch size greater than one, which is not feasible when processing a single large image or in scenarios with limited batch capacity.
\end{itemize}

For these reasons, we prioritize the interweaved parallelism approach described earlier, which retains the benefits of asynchronous scheduling while requiring fewer buffers and achieving higher GPU utilization.

\begin{table}[htbp]
  \centering
  \resizebox{\columnwidth}{!}{
      \begin{tabular}{ll*{4}{c}}
        \toprule
        \multirow{2}{*}{Model} & \multirow{2}{*}{GPUs} & \multicolumn{4}{c}{Batch Size} \\
        \cmidrule(lr){3-6}
        & & 4 & 8 & 16 & 32 \\
        \midrule
        \multirow{2}{*}{DiT-MoE-XL} 
        & 4 & 62.9\% & 70.9\% & 73.8\% & 74.4\% \\
        & 8 &75.6\% & 78.1\% & 79.0\% & 79.2\% \\
        \addlinespace
        \multirow{2}{*}{DiT-MoE-G}
        & 4 & 50.7\% & 56.8\% & 59.3\% & 61.1\% \\
        & 8 & 64.7\% & 67.8\% & 69.2\% & 68.9\% \\
        \bottomrule
      \end{tabular}
    }
  \caption{All-to-All communication time percentage in synchronous expert parallelism.}
  \label{tab:all2all_InEP}
\end{table}



\section{More Implementation Details}
\label{sec:more-impl}
\noindent\textbf{Setup.} The models used in our study are the publicly available versions of DiT-MoE from Huggingface. In our experimental results, the batch size refers to the local batch size, representing the number of samples processed per device. All asynchronous methods apply the same number of synchronized steps(e.g. warmup) post-start.

\noindent\textbf{Expert Score Scaling.} There are two approaches for scaling results after expert processing: (1) using the latest router scores computed in the current step, which provides fresher scores, and (2) using the router scores corresponding to the stale expert input, offering better alignment with the activations used. The selection of the scores has little impact on performance. For fairness, both displaced parallelism and \dice{} use the stale router scores for scaling.

\noindent\textbf{Extending Image Sizes.} The public DiT-MoE model supports relatively low resolution image dimensions. To extend our experiments to larger images, we initialize positional embeddings for other sizes. Although this adjustment prevents the model from generating meaningful images, it enables us to evaluate latency, memory usage, and speedup across different resolutions.

\begin{algorithm}[t]
\caption{Interweaved Parallelism}\label{alg:interweaved}
\For{each layer}{
  $x \gets \text{attn}(x)$ \\  
  $[s,\, idx] \gets \text{router}(x)$ \\
  $h_d \gets \text{async\_dispatch}(x, idx)$ \tcp{Launch non-blocking dispatch for current layer}  
  $\text{prev\_layer\_x}_d \gets \text{wait}(\text{prev\_layer\_h}_d)$ \tcp{Wait for previous layer's dispatch result from current step}  
  $\text{prev\_layer\_x}_e \gets \text{prev\_layer\_expert}(\text{prev\_layer\_x}_d)$ \tcp{Process tokens using previous layer's experts}
  $\text{prev\_layer\_h}_c \gets \text{async\_combine}(\text{prev\_layer\_x}_e)$ \tcp{Launch non-blocking combine for previous layer's outputs}  
  $\text{x}_c \gets \text{wait}(h_c^{\text{prev}})$ \tcp{Wait for previous step's combine result for current layer}  
  $x \gets \text{scale}(\text{x}_c, s)$ \tcp{Scale outputs using router scores}
}
\end{algorithm}

\noindent\textbf{Interweaved Parallelism} restructures the execution flow, effectively ``folding'' it. It processes expert computations within the same step while maintaining asynchronous communication, as outlined in \cref{alg:interweaved}.

\noindent\textbf{Conditional Communication}
For each token-expert pair \((i,e)\) at step \(t\), conditional communication dynamically decides whether to send fresh activations or reuse cached results. 
As shown in Algorithm~\ref{alg:cond}, the top-1 expert for each token always receives fresh data, ensuring critical tokens remain up-to-date. 
Lower-priority experts (i.e., those not ranked top-1) reuse their previous activations most of the time and only receive an update every \(n\) steps. 
This design leverages the weighted-sum mechanic of MoE to maintain high-impact tokens' freshness while reducing communication overhead for less critical tokens.
\begin{algorithm}[t]
\caption{Conditional Communication}\label{alg:cond}
\For{token-expert pair $(i,e)$ at step $t$}{
    \If{$e$ is top-1 expert}{
        Transmit Token \tcp*{Always fresh}
    }
    \Else{
        \eIf{$t \bmod n \neq 0$}{ 
            Reuse stale $\mathbf{h}_i^e$\; 
        }{
            Transmit Token\; \tcp*{Update every $n$ steps}
        }
    }
}
\end{algorithm}

\section{Discussion}

\noindent\textbf{Limitations.} Although \dice{} demonstrates significant gains in inference efficiency, there remain avenues for further improvements. Optimized kernels and more efficient NCCL operations could help further reduce latency. Additionally, integrating \dice{} with existing expert parallelism optimizations offers opportunities to enhance its efficiency and scalability. Another limitation lies in the availability of MoE-based diffusion models, which restricts our evaluations to a limited set of configurations. As more MoE-based diffusion models are developed, \dice{} can be validated and refined across a broader range of scenarios.

\noindent\textbf{Influence of Shared Experts.} 
The architecture of DiT-MoE includes shared experts, a proven mechanism for enhancing MoE performance. We hypothesize that these shared experts may help mitigate the impact of staleness in similarity-based asynchronous parallelism. Unlike routed experts, whose outputs can become stale, the shared expert's computations are always up-to-date (as they are duplicated across devices), potentially providing fresh information to balance the delayed outputs from routed experts. This characteristic might play a role in the performance of \dice{}, particularly when compared to DistriFusion. While both approaches exhibit a staleness of 1, \dice{} confines staleness to routed experts while benefiting from the shared expert's fresh contributions. This suggests a possible advantage for \dice{} in MoE-based models.

\noindent\textbf{Applicability to NVLink.} While our experiments are conducted on PCIe-connected GPUs, \dice{} remains applicable when MoE models are served under NVLink and InfiniBand-based multi-node deployments \cite{interlayer}. In such settings, all-to-all communication can contribute up to \textbf{76\%} of total inference latency~\cite{interlayer}, suggesting that \dice{} might offer even greater benefits in these environments.

\noindent\textbf{Integration with Existing Expert Parallelism Optimizations.} Existing expert parallelism optimizations—such as expert shadowing~\cite{fastermoe}, topology-aware routing~\cite{fastermoe, locmoe}, and affinity-based methods~\cite{interlayer}—address orthogonal challenges; respectively: load balancing, network topology, and expert placement. These techniques are \textbf{potentially integrable with \dice{}} rather than alternatives, and combining them could further enhance overall efficiency.

\noindent\textbf{Usage of Router Similarity.} Routers assign token destinations during all-to-all communication, and their inherent redundancy is crucial to maintain consistent token-to-expert assignments. Without this similarity, asynchronous (displaced) operations would disrupt token-expert assignments and degrade performance.

\section{Additional Experiments}
\noindent\textbf{All-to-All Blocking Latency in Expert Parallelism.}
We measure all-to-all latency on the same hardware as Section \ref{sec:experiments}. \Cref{tab:all2all_InEP} reveals a significant bottleneck inherent in synchronous Expert Parallelism for DiT-MoE models: the all-to-all communication overhead. The experimental data indicates that, across all tested configurations, all-to-all communication time significantly surpasses half of the total inference time. Furthermore, a pronounced upward trend is observed in this percentage as the batch size increases, reaching a strikingly high 79.2\%. This underscores the necessity of our proposed approach to mitigate this communication bottleneck and achieve substantial inference acceleration.

\begin{figure*}[h]
    \centering
    \vspace{10pt}
    \includegraphics[width=\textwidth]{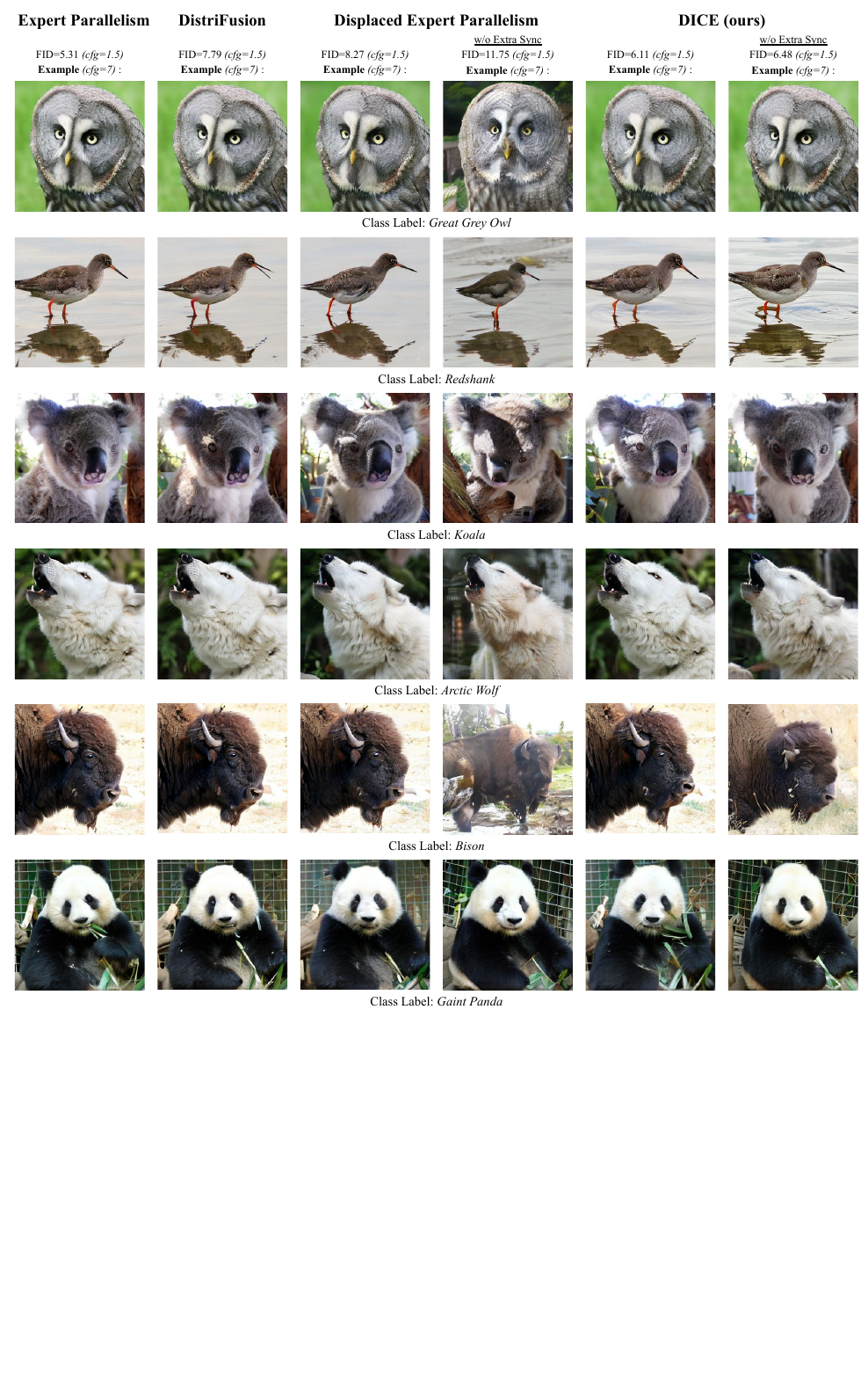}
    \vspace{-20pt}
    \caption{Extra qualitative results.}
    \label{fig:sample_sup}
\end{figure*}

\noindent\textbf{Quality results.} In Figure \ref{fig:sample_sup}, we present additional qualitative results across six different classes. Notably, only the displaced expert parallelism without extra synchronization demonstrates considerable deviations. In particular, the images produced by \dice{} closely resemble the synced version.



\clearpage

%% file: main.bib
@String(CVPR= {IEEE Conf. Comput. Vis. Pattern Recog.})

@String(ICCV= {Int. Conf. Comput. Vis.})

@String(NIPS= {Adv. Neural Inform. Process. Syst.})

@String(CVPR  = {CVPR})

@String(ICCV  = {ICCV})

@String(NIPS  = {NeurIPS})

@article{ddpm,
  title={Denoising diffusion probabilistic models},
  author={Ho, Jonathan and Jain, Ajay and Abbeel, Pieter},
  journal={Advances in neural information processing systems},
  volume={33},
  pages={6840--6851},
  year={2020}
}

@article{dhariwal2021diffusion,
  title={Diffusion models beat gans on image synthesis},
  author={Dhariwal, Prafulla and Nichol, Alexander},
  journal={Advances in neural information processing systems},
  volume={34},
  pages={8780--8794},
  year={2021}
}

@article{restoration,
  title={Denoising diffusion restoration models},
  author={Kawar, Bahjat and Elad, Michael and Ermon, Stefano and Song, Jiaming},
  journal={Advances in Neural Information Processing Systems},
  volume={35},
  pages={23593--23606},
  year={2022}
}

@article{score,
  title={Score-based generative modeling in latent space},
  author={Vahdat, Arash and Kreis, Karsten and Kautz, Jan},
  journal={Advances in neural information processing systems},
  volume={34},
  pages={11287--11302},
  year={2021}
}

@article{video,
  title={Imagen video: High definition video generation with diffusion models},
  author={Ho, Jonathan and Chan, William and Saharia, Chitwan and Whang, Jay and Gao, Ruiqi and Gritsenko, Alexey and Kingma, Diederik P and Poole, Ben and Norouzi, Mohammad and Fleet, David J and others},
  journal={arXiv preprint arXiv:2210.02303},
  year={2022}
}

@article{luo2023videofusion,
  title={Videofusion: Decomposed diffusion models for high-quality video generation},
  author={Luo, Zhengxiong and Chen, Dayou and Zhang, Yingya and Huang, Yan and Wang, Liang and Shen, Yujun and Zhao, Deli and Zhou, Jingren and Tan, Tieniu},
  journal={arXiv preprint arXiv:2303.08320},
  year={2023}
}

@article{ditmoe,
  title={Scaling Diffusion Transformers to 16 Billion Parameters},
  author={Fei, Zhengcong and Fan, Mingyuan and Yu, Changqian and Li, Debang and Huang, Junshi},
  journal={arXiv preprint arXiv:2407.11633},
  year={2024}
}

@InProceedings{dit,
    author    = {Peebles, William and Xie, Saining},
    title     = {Scalable Diffusion Models with Transformers},
    booktitle = {Proceedings of the IEEE/CVF International Conference on Computer Vision (ICCV)},
    month     = {October},
    year      = {2023},
    pages     = {4195-4205}
}

@article{lu2024tmc,
title={Data-Aware Gradient Compression for FL in Communication-Constrained Mobile Computing},
author={Lu, Rongwei and Jiang, Yutong and Mao, Yinan and Tang, Chen and Chen, Bin and Cui, Laizhong and Wang, Zhi},
journal={IEEE Transactions on Mobile Computing},
year={2024},
publisher={IEEE}
}

@inproceedings{lu2023dagc,
title={Dagc: Data-aware adaptive gradient compression},
author={Lu, Rongwei and Song, Jiajun and Chen, Bin and Cui, Laizhong and Wang, Zhi},
booktitle={IEEE INFOCOM 2023-IEEE Conference on Computer Communications},
pages={1--10},
year={2023},
organization={IEEE}
}

@misc{moe,
	title = {Outrageously {Large} {Neural} {Networks}: {The} {Sparsely}-{Gated} {Mixture}-of-{Experts} {Layer}},
	shorttitle = {Outrageously {Large} {Neural} {Networks}},
	url = {http://arxiv.org/abs/1701.06538},
	doi = {10.48550/arXiv.1701.06538},
	urldate = {2023-10-24},
	publisher = {arXiv},
	author = {Shazeer, Noam and Mirhoseini, Azalia and Maziarz, Krzysztof and Davis, Andy and Le, Quoc and Hinton, Geoffrey and Dean, Jeff},
	month = jan,
	year = {2017},
	note = {arXiv:1701.06538 [cs, stat]},
}

@misc{megatron,
	title = {Megatron-{LM}: {Training} {Multi}-{Billion} {Parameter} {Language} {Models} {Using} {Model} {Parallelism}},
	shorttitle = {Megatron-{LM}},
	url = {http://arxiv.org/abs/1909.08053},
	doi = {10.48550/arXiv.1909.08053},
	urldate = {2023-10-29},
	publisher = {arXiv},
	author = {Shoeybi, Mohammad and Patwary, Mostofa and Puri, Raul and LeGresley, Patrick and Casper, Jared and Catanzaro, Bryan},
	month = mar,
	year = {2020},
	note = {arXiv:1909.08053 [cs]},
}

@misc{gshard,
	title = {{GShard}: {Scaling} {Giant} {Models} with {Conditional} {Computation} and {Automatic} {Sharding}},
	shorttitle = {{GShard}},
	url = {http://arxiv.org/abs/2006.16668},
	doi = {10.48550/arXiv.2006.16668},
	urldate = {2023-10-29},
	publisher = {arXiv},
	author = {Lepikhin, Dmitry and Lee, HyoukJoong and Xu, Yuanzhong and Chen, Dehao and Firat, Orhan and Huang, Yanping and Krikun, Maxim and Shazeer, Noam and Chen, Zhifeng},
	month = jun,
	year = {2020},
	note = {arXiv:2006.16668 [cs, stat]},
}

@inproceedings{fastermoe,
    author = {He, Jiaao and Zhai, Jidong and Antunes, Tiago and Wang, Haojie and Luo, Fuwen and Shi, Shangfeng and Li, Qin},
    title = {FasterMoE: modeling and optimizing training of large-scale dynamic pre-trained models},
    year = {2022},
    isbn = {9781450392044},
    publisher = {Association for Computing Machinery},
    address = {New York, NY, USA},
    url = {https://doi.org/10.1145/3503221.3508418},
    doi = {10.1145/3503221.3508418},
    booktitle = {Proceedings of the 27th ACM SIGPLAN Symposium on Principles and Practice of Parallel Programming},
    pages = {120–134},
    numpages = {15},
    location = {Seoul, Republic of Korea},
    series = {PPoPP '22}
}

@InProceedings{baselayers,
  title = 	 {BASE Layers: Simplifying Training of Large, Sparse Models},
  author =       {Lewis, Mike and Bhosale, Shruti and Dettmers, Tim and Goyal, Naman and Zettlemoyer, Luke},
  booktitle = 	 {Proceedings of the 38th International Conference on Machine Learning},
  pages = 	 {6265--6274},
  year = 	 {2021},
  editor = 	 {Meila, Marina and Zhang, Tong},
  volume = 	 {139},
  series = 	 {Proceedings of Machine Learning Research},
  month = 	 {18--24 Jul},
  publisher =    {PMLR},
  url = 	 {https://proceedings.mlr.press/v139/lewis21a.html},
}

@misc{fastmoe,
	title = {{FastMoE}: {A} {Fast} {Mixture}-of-{Expert} {Training} {System}},
	shorttitle = {{FastMoE}},
	url = {http://arxiv.org/abs/2103.13262},
	doi = {10.48550/arXiv.2103.13262},
	urldate = {2023-11-04},
	publisher = {arXiv},
	author = {He, Jiaao and Qiu, Jiezhong and Zeng, Aohan and Yang, Zhilin and Zhai, Jidong and Tang, Jie},
	month = mar,
	year = {2021},
	note = {arXiv:2103.13262 [cs]},
}

@InProceedings{deepspeedmoe,
  title = 	 {{D}eep{S}peed-{M}o{E}: Advancing Mixture-of-Experts Inference and Training to Power Next-Generation {AI} Scale},
  author =       {Rajbhandari, Samyam and Li, Conglong and Yao, Zhewei and Zhang, Minjia and Aminabadi, Reza Yazdani and Awan, Ammar Ahmad and Rasley, Jeff and He, Yuxiong},
  booktitle = 	 {Proceedings of the 39th International Conference on Machine Learning},
  pages = 	 {18332--18346},
  year = 	 {2022},
  editor = 	 {Chaudhuri, Kamalika and Jegelka, Stefanie and Song, Le and Szepesvari, Csaba and Niu, Gang and Sabato, Sivan},
  volume = 	 {162},
  series = 	 {Proceedings of Machine Learning Research},
  month = 	 {17--23 Jul},
  publisher =    {PMLR},
  url = 	 {https://proceedings.mlr.press/v162/rajbhandari22a.html},
}

@misc{seqpara,
	title = {Sequence {Parallelism}: {Long} {Sequence} {Training} from {System} {Perspective}},
	shorttitle = {Sequence {Parallelism}},
	url = {http://arxiv.org/abs/2105.13120},
	doi = {10.48550/arXiv.2105.13120},
	urldate = {2024-01-10},
	publisher = {arXiv},
	author = {Li, Shenggui and Xue, Fuzhao and Baranwal, Chaitanya and Li, Yongbin and You, Yang},
	month = may,
	year = {2022},
	note = {arXiv:2105.13120 [cs]},
}

@inproceedings {lina,
    author = {Jiamin Li and Yimin Jiang and Yibo Zhu and Cong Wang and Hong Xu},
    title = {Accelerating Distributed {MoE} Training and Inference with Lina},
    booktitle = {2023 USENIX Annual Technical Conference (USENIX ATC 23)},
    year = {2023},
    isbn = {978-1-939133-35-9},
    address = {Boston, MA},
    pages = {945--959},
    url = {https://www.usenix.org/conference/atc23/presentation/li-jiamin},
    publisher = {USENIX Association},
    month = jul
}

@misc{pipefusion,
	title = {{PipeFusion}: {Displaced} {Patch} {Pipeline} {Parallelism} for {Inference} of {Diffusion} {Transformer} {Models}},
	shorttitle = {{PipeFusion}},
	url = {http://arxiv.org/abs/2405.14430},
	doi = {10.48550/arXiv.2405.14430},
	urldate = {2024-09-03},
	publisher = {arXiv},
	author = {Wang, Jiannan and Fang, Jiarui and Li, Aoyu and Yang, PengCheng},
	month = may,
	year = {2024},
	note = {arXiv:2405.14430 [cs]},
}

@misc{learn2cache,
	title = {Learning-to-{Cache}: {Accelerating} {Diffusion} {Transformer} via {Layer} {Caching}},
	shorttitle = {Learning-to-{Cache}},
	url = {http://arxiv.org/abs/2406.01733},
	doi = {10.48550/arXiv.2406.01733},
	urldate = {2024-09-03},
	publisher = {arXiv},
	author = {Ma, Xinyin and Fang, Gongfan and Mi, Michael Bi and Wang, Xinchao},
	month = jun,
	year = {2024},
	note = {arXiv:2406.01733 [cs]},
}

@InProceedings{deepcache,
    author    = {Ma, Xinyin and Fang, Gongfan and Wang, Xinchao},
    title     = {DeepCache: Accelerating Diffusion Models for Free},
    booktitle = {Proceedings of the IEEE/CVF Conference on Computer Vision and Pattern Recognition (CVPR)},
    month     = {June},
    year      = {2024},
    pages     = {15762-15772}
}

@misc{asyncdiff,
	title = {{AsyncDiff}: {Parallelizing} {Diffusion} {Models} by {Asynchronous} {Denoising}},
	shorttitle = {{AsyncDiff}},
	url = {http://arxiv.org/abs/2406.06911},
	language = {en},
	urldate = {2024-09-18},
	publisher = {arXiv},
	author = {Chen, Zigeng and Ma, Xinyin and Fang, Gongfan and Tan, Zhenxiong and Wang, Xinchao},
	month = jun,
	year = {2024},
	note = {arXiv:2406.06911 [cs]},
}

@misc{fora,
	title = {{FORA}: {Fast}-{Forward} {Caching} in {Diffusion} {Transformer} {Acceleration}},
	shorttitle = {{FORA}},
	url = {http://arxiv.org/abs/2407.01425},
	language = {en},
	urldate = {2024-09-18},
	publisher = {arXiv},
	author = {Selvaraju, Pratheba and Ding, Tianyu and Chen, Tianyi and Zharkov, Ilya and Liang, Luming},
	month = jul,
	year = {2024},
	note = {arXiv:2407.01425 [cs]},
}

@INPROCEEDINGS {interlayer,
author = { Yao, Jinghan and Anthony, Quentin and Shafi, Aamir and Subramoni, Hari and DK Panda, Dhabaleswar K. },
booktitle = { 2024 IEEE International Parallel and Distributed Processing Symposium (IPDPS) },
title = {{ Exploiting Inter-Layer Expert Affinity for Accelerating Mixture-of-Experts Model Inference }},
year = {2024},
volume = {},
ISSN = {},
pages = {915-925},
doi = {10.1109/IPDPS57955.2024.00086},
url = {https://doi.ieeecomputersociety.org/10.1109/IPDPS57955.2024.00086},
publisher = {IEEE Computer Society},
address = {Los Alamitos, CA, USA},
month =May}

@inproceedings{distri,
	address = {Seattle, WA, USA},
	title = {{DistriFusion}: {Distributed} {Parallel} {Inference} for {High}-{Resolution} {Diffusion} {Models}},
	copyright = {https://doi.org/10.15223/policy-029},
	isbn = {9798350353006},
	shorttitle = {{DistriFusion}},
	url = {https://ieeexplore.ieee.org/document/10657352/},
	doi = {10.1109/CVPR52733.2024.00686},
	language = {en},
	urldate = {2024-10-30},
	booktitle = {2024 {IEEE}/{CVF} {Conference} on {Computer} {Vision} and {Pattern} {Recognition} ({CVPR})},
	publisher = {IEEE},
	author = {Li, Muyang and Cai, Tianle and Cao, Jiaxin and Zhang, Qinsheng and Cai, Han and Bai, Junjie and Jia, Yangqing and Li, Kai and Han, Song},
	month = jun,
	year = {2024},
	pages = {7183--7193},
}

@inproceedings{transformer,
     author = {Vaswani, Ashish and Shazeer, Noam and Parmar, Niki and Uszkoreit, Jakob and Jones, Llion and Gomez, Aidan N and Kaiser, \L ukasz and Polosukhin, Illia},
     booktitle = {Advances in Neural Information Processing Systems},
     editor = {I. Guyon and U. Von Luxburg and S. Bengio and H. Wallach and R. Fergus and S. Vishwanathan and R. Garnett},
     pages = {},
     publisher = {Curran Associates, Inc.},
     title = {Attention is All you Need},
     url = {https://proceedings.neurips.cc/paper_files/paper/2017/file/3f5ee243547dee91fbd053c1c4a845aa-Paper.pdf},
     volume = {30},
     year = {2017}
}

@misc{rf,
      title={Flow Straight and Fast: Learning to Generate and Transfer Data with Rectified Flow}, 
      author={Xingchao Liu and Chengyue Gong and Qiang Liu},
      year={2022},
      eprint={2209.03003},
      archivePrefix={arXiv},
      primaryClass={cs.LG},
      url={https://arxiv.org/abs/2209.03003}, 
}

@misc{cfg,
      title={Classifier-Free Diffusion Guidance}, 
      author={Jonathan Ho and Tim Salimans},
      year={2022},
      eprint={2207.12598},
      archivePrefix={arXiv},
      primaryClass={cs.LG},
      url={https://arxiv.org/abs/2207.12598}, 
}

@inproceedings{imagenet,
  title={Imagenet: A large-scale hierarchical image database},
  author={Deng, Jia and Dong, Wei and Socher, Richard and Li, Li-Jia and Li, Kai and Fei-Fei, Li},
  booktitle={2009 IEEE conference on computer vision and pattern recognition},
  pages={248--255},
  year={2009},
  organization={Ieee}
}

@misc{locmoe,
      title={LocMoE: A Low-Overhead MoE for Large Language Model Training}, 
      author={Jing Li and Zhijie Sun and Xuan He and Li Zeng and Yi Lin and Entong Li and Binfan Zheng and Rongqian Zhao and Xin Chen},
      year={2024},
      eprint={2401.13920},
      archivePrefix={arXiv},
      primaryClass={cs.LG},
      url={https://arxiv.org/abs/2401.13920}, 
}

@misc{ditmoe-git,
	title = {feizc/{DiT}-{MoE}},
	url = {https://github.com/feizc/DiT-MoE},
	urldate = {2024-10-30},
	author = {Zhengcong Fei},
	month = oct,
	year = {2024},
	note = {original-date: 2024-06-25T08:27:55Z},
}

@incollection{torch,
    title = {PyTorch: An Imperative Style, High-Performance Deep Learning Library},
    author = {Paszke, Adam and Gross, Sam and Massa, Francisco and Lerer, Adam and Bradbury, James and Chanan, Gregory and Killeen, Trevor and Lin, Zeming and Gimelshein, Natalia and Antiga, Luca and Desmaison, Alban and Kopf, Andreas and Yang, Edward and DeVito, Zachary and Raison, Martin and Tejani, Alykhan and Chilamkurthy, Sasank and Steiner, Benoit and Fang, Lu and Bai, Junjie and Chintala, Soumith},
    booktitle = {Advances in Neural Information Processing Systems 32},
    pages = {8024--8035},
    year = {2019},
    publisher = {Curran Associates, Inc.},
    url = {http://papers.neurips.cc/paper/9015-pytorch-an-imperative-style-high-performance-deep-learning-library.pdf}
}

@inproceedings{fid,
 author = {Heusel, Martin and Ramsauer, Hubert and Unterthiner, Thomas and Nessler, Bernhard and Hochreiter, Sepp},
 booktitle = {Advances in Neural Information Processing Systems},
 editor = {I. Guyon and U. Von Luxburg and S. Bengio and H. Wallach and R. Fergus and S. Vishwanathan and R. Garnett},
 pages = {},
 publisher = {Curran Associates, Inc.},
 title = {GANs Trained by a Two Time-Scale Update Rule Converge to a Local Nash Equilibrium},
 url = {https://proceedings.neurips.cc/paper_files/paper/2017/file/8a1d694707eb0fefe65871369074926d-Paper.pdf},
 volume = {30},
 year = {2017}
}

@inproceedings{is,
 author = {Salimans, Tim and Goodfellow, Ian and Zaremba, Wojciech and Cheung, Vicki and Radford, Alec and Chen, Xi and Chen, Xi},
 booktitle = {Advances in Neural Information Processing Systems},
 editor = {D. Lee and M. Sugiyama and U. Luxburg and I. Guyon and R. Garnett},
 pages = {},
 publisher = {Curran Associates, Inc.},
 title = {Improved Techniques for Training GANs},
 url = {https://proceedings.neurips.cc/paper_files/paper/2016/file/8a3363abe792db2d8761d6403605aeb7-Paper.pdf},
 volume = {29},
 year = {2016}
}

@article{sfid,
  title={Generating Images with Sparse Representations},
  author={Charlie Nash and Jacob Menick and Sander Dieleman and Peter W. Battaglia},
  journal={ArXiv},
  year={2021},
  volume={abs/2103.03841},
  url={https://api.semanticscholar.org/CorpusID:232135095}
}

@inproceedings{prec-recall,
 author = {Kynk\"{a}\"{a}nniemi, Tuomas and Karras, Tero and Laine, Samuli and Lehtinen, Jaakko and Aila, Timo},
 booktitle = {Advances in Neural Information Processing Systems},
 editor = {H. Wallach and H. Larochelle and A. Beygelzimer and F. d\textquotesingle Alch\'{e}-Buc and E. Fox and R. Garnett},
 pages = {},
 publisher = {Curran Associates, Inc.},
 title = {Improved Precision and Recall Metric for Assessing Generative Models},
 url = {https://proceedings.neurips.cc/paper_files/paper/2019/file/0234c510bc6d908b28c70ff313743079-Paper.pdf},
 volume = {32},
 year = {2019}
}

@misc{ditmoe-hf,
        author = {feizhengcong},
	title = {feizhengcong/{DiT}-{MoE} · {Hugging} {Face}},
	url = {https://huggingface.co/feizhengcong/DiT-MoE},
	urldate = {2024-11-10},
        year = {2024}
}

@inproceedings{wimbauer2024cache,
  title={Cache me if you can: Accelerating diffusion models through block caching},
  author={Wimbauer, Felix and Wu, Bichen and Schoenfeld, Edgar and Dai, Xiaoliang and Hou, Ji and He, Zijian and Sanakoyeu, Artsiom and Zhang, Peizhao and Tsai, Sam and Kohler, Jonas and others},
  booktitle={Proceedings of the IEEE/CVF Conference on Computer Vision and Pattern Recognition},
  pages={6211--6220},
  year={2024}
}

@inproceedings{habibian2024clockwork,
  title={Clockwork Diffusion: Efficient Generation With Model-Step Distillation},
  author={Habibian, Amirhossein and Ghodrati, Amir and Fathima, Noor and Sautiere, Guillaume and Garrepalli, Risheek and Porikli, Fatih and Petersen, Jens},
  booktitle={Proceedings of the IEEE/CVF Conference on Computer Vision and Pattern Recognition},
  pages={8352--8361},
  year={2024}
}

@article{so2023frdiff,
  title={FRDiff: Feature Reuse for Universal Training-free Acceleration of Diffusion Models},
  author={So, Junhyuk and Lee, Jungwon and Park, Eunhyeok},
  journal={arXiv preprint arXiv:2312.03517},
  year={2023}
}

@inproceedings{rap,
author = {Xue, Zeyue and Song, Guanglu and Guo, Qiushan and Liu, Boxiao and Zong, Zhuofan and Liu, Yu and Luo, Ping},
title = {RAPHAEL: text-to-image generation via large mixture of diffusion Paths},
year = {2023},
publisher = {Curran Associates Inc.},
address = {Red Hook, NY, USA},
booktitle = {Proceedings of the 37th International Conference on Neural Information Processing Systems},
articleno = {1806},
numpages = {14},
location = {New Orleans, LA, USA},
series = {NIPS '23}
}

@InProceedings{qdit,
    author    = {Chen, Lei and Meng, Yuan and Tang, Chen and Ma, Xinzhu and Jiang, Jingyan and Wang, Xin and Wang, Zhi and Zhu, Wenwu},
    title     = {Q-DiT: Accurate Post-Training Quantization for Diffusion Transformers},
    booktitle = {Proceedings of the Computer Vision and Pattern Recognition Conference (CVPR)},
    month     = {June},
    year      = {2025},
    pages     = {28306-28315}
}

@misc{dist_survey,
      title={Beyond A Single AI Cluster: A Survey of Decentralized LLM Training}, 
      author={Haotian Dong and Jingyan Jiang and Rongwei Lu and Jiajun Luo and Jiajun Song and Bowen Li and Ying Shen and Zhi Wang},
      year={2025},
      eprint={2503.11023},
      archivePrefix={arXiv},
      primaryClass={cs.DC},
      url={https://arxiv.org/abs/2503.11023}, 
}
